%% file: 2026_arXiv_manuscript.tex
\documentclass{article}
\usepackage{./00_template/arxiv}
\usepackage[utf8]{inputenc}
\usepackage[T1]{fontenc}
\usepackage[
  colorlinks=true,
  linkcolor=black,
  citecolor=blue,
  urlcolor=blue,
]{hyperref}
\usepackage{caption}
\DeclareCaptionFont{sevenpt}{\fontsize{9pt}{12pt}\selectfont}
\usepackage{url}
\usepackage{nicefrac}
\usepackage{microtype}
\usepackage[round]{natbib}
\usepackage{doi}
\usepackage{amsmath,amssymb,amsfonts,amsthm,mathrsfs}
\usepackage[scr=boondoxo]{mathalfa}
\usepackage{mathtools}
\usepackage{dsfont}
\usepackage{euscript}
\newcommand{\tr}{\operatorname{tr}}
\newcommand{\rank}{\operatorname{rank}}
\newcommand{\diag}{\operatorname{diag}}

\newcommand{\expect}{\operatorname{\mathbb{E}}}

\newcommand{\argmax}{\operatorname{arg\,max}}

\usepackage{algorithmic}
\usepackage{graphicx}
\usepackage{textcomp}
\usepackage{xcolor}
\usepackage{stfloats}
\usepackage{enumitem}
\usepackage{subcaption}
\usepackage{rotating}
\usepackage{colortbl}
\usepackage{xfrac}
\usepackage{tabularx}
\usepackage{array}
\usepackage{booktabs}
\usepackage{threeparttable}
\usepackage{makecell}
\usepackage{multicol}
\usepackage{multirow}
\usepackage{listingsutf8}
\usepackage{float}
\usepackage{siunitx}
\usepackage{textcomp}
\usepackage{blindtext}
\usepackage{lipsum}
\usepackage{soul}
\usepackage{printlen}
\uselengthunit{cm}
\newsavebox{\mycaptionbox}
\title{Bayesian Calibration of Effective Material Parameters in Electric Machine Stators}
\date{} 					
\newif\ifanonymous
\anonymousfalse 
\ifanonymous
  \vspace{3cm}
\else
\author{ \href{https://orcid.org/0009-0007-6208-1888}{\includegraphics[scale=0.06]{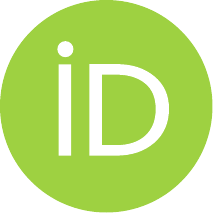}\hspace{1mm}Patrick D.~Gs\"ols\textsuperscript{$\star$}} \\
	Silicon Austria Labs GmbH\\
	Graz University of Technology\\
	8010 Graz, Austria\\
	\And
	\href{https://orcid.org/0009-0005-6967-3515}{\includegraphics[scale=0.06]{./01_figures/orcid.pdf}\hspace{1mm}Christian~M.~Riener} \\
	Silicon Austria Labs GmbH\\
	8010 Graz, Austria\\
	\And
	\href{https://orcid.org/0009-0002-2180-2793}{\includegraphics[scale=0.06]{./01_figures/orcid.pdf}\hspace{1mm}Markus~Flicker} \\
	Silicon Austria Labs GmbH\\
	8010 Graz, Austria\\
	\And
	\href{https://orcid.org/0000-0001-7641-104X}{\includegraphics[scale=0.06]{./01_figures/orcid.pdf}\hspace{1mm}Jose~Romero~Lopera} \\
	Silicon Austria Labs GmbH\\
	8010 Graz, Austria\\
	\And
	\href{https://orcid.org/0009-0003-0755-7960}{\includegraphics[scale=0.06]{./01_figures/orcid.pdf}\hspace{1mm}Jan~P.~Henninger\textsuperscript{$\star$}} \\
	Graz University of Technology\\
	8010 Graz, Austria\\
    \And
	\href{https://orcid.org/0000-0002-4738-9602}{\includegraphics[scale=0.06]{./01_figures/orcid.pdf}\hspace{1mm}Bibhu~Prasad~Nayak\textsuperscript{$\star$}} \\
	Graz University of Technology\\
	8010 Graz, Austria\\
	\And
	\href{https://orcid.org/0000-0000-0000-0000}{\includegraphics[scale=0.06]{./01_figures/orcid.pdf}\hspace{1mm}Jan~Biermann} \\
	BMW Group\\
	80788 Munich, Germany\\
	\And
	\href{https://orcid.org/0000-0002-9244-6426}{\includegraphics[scale=0.06]{./01_figures/orcid.pdf}\hspace{1mm}Bernhard~Auinger} \\
	Silicon Austria Labs GmbH\\
	8010 Graz, Austria\\
	\And
	\href{https://orcid.org/0000-0002-4650-9094}{\includegraphics[scale=0.06]{./01_figures/orcid.pdf}\hspace{1mm}Jan~C.~Hansen\textsuperscript{$\star$}} \\
	Silicon Austria Labs GmbH\\
	Graz University of Technology\\
	8010 Graz, Austria\\
}
\fi
\renewcommand{\headeright}{A Preprint}
\renewcommand{\undertitle}{A Preprint --- Non-Peer Reviewed}
\renewcommand{\shorttitle}{}
\ifanonymous
\hypersetup{
pdftitle={Bayesian Calibration of Effective Material Parameters in Electric Machine Stators},
pdfkeywords={Electromagnetic compatibility, Bayesian Inference, Bayesian Model Calibration, Gaussian Processes, Inverse problems, Magnetic materials, Monte Carlo methods, Traction motors, Wideband Kriging},
}
\else
\hypersetup{
pdftitle={Bayesian Calibration of Effective Material Parameters in Electric Machine Stators},
pdfsubject={A Preprint --- Non-Peer Reviewed arXiv Version},
pdfauthor={Patrick~D.~Gs\"ols, Christian~M.~Riener, Markus~Flicker, Jose~Romero~Lopera, Jan~P.~Henninger, Bibhu~Prasad~Nayak, Jan~Biermann, Bernhard~Auinger, Jan~C.~Hansen},
pdfkeywords={Electromagnetic compatibility, Bayesian Inference, Bayesian Model Calibration, Gaussian Process Regression, Inverse problems, Magnetic materials, Monte Carlo methods, Traction motors, Wideband Kriging},
}
\fi
\newcommand{\figref}[1]{Fig.~\ref{#1}}
\newcommand{\tabref}[1]{Table~\ref{#1}}
\newcommand{\secref}[1]{Section~\ref{#1}}

\usepackage{nomencl}
\makenomenclature
\input{02_nomenclature_and_acronyms/01_nomenclature.tex}
\usepackage[acronym]{glossaries}
\makeglossaries
\input{02_nomenclature_and_acronyms/02_acronyms.tex}
\begin{document}
\maketitle
\vspace{-3.5\baselineskip}
\ifanonymous
\begin{scriptsize}
\begin{center}
 \vspace{5cm}
\end{center}
\end{scriptsize}
\else
\begin{scriptsize}
\begin{center}
\textsuperscript{$\star$}Members of the Christian Doppler Laboratory for EMC Aware Robust Electronic Systems at the Institute of Electronics,\\Graz University of Technology, 8010 Graz, Austria\\\vspace{10pt}
\textsc{Corresponding Author}: Patrick D. Gs\"ols (e-mail: \href{mailto:patrick.gsoels@silicon-austria.com;patrick.gsoels@student.tugraz.at}{\texttt{patrick.gsoels@\{silicon-austria.com,student.tugraz.at\}}})
\end{center}
\end{scriptsize}
\fi
\vspace{12pt}
%
%
\begin{abstract}
  \input{03_abstract/01_abstract.tex}
\end{abstract}
%
%
%
%
%
%
%
%
%
\input{04_main_content/01_introduction.tex}
\input{04_main_content/02_wideband_calibration_problem.tex}
\input{04_main_content/03_bayesian_inference_formulation.tex}
\input{04_main_content/04_bayesian_inference_results.tex}
\input{04_main_content/05_conclusion.tex}
%
%
%
\appendix
%
\input{05_appendix/01_appendix.tex}
%
%
%
%
\ifanonymous
\begin{center}
 \vspace{0pt}
\end{center}
\else
\phantomsection
\addcontentsline{toc}{section*}{Acknowledgement}
\input{06_acknowledgement/01_acknowledgement.tex}
\fi
%
%
\phantomsection
\addcontentsline{toc}{section*}{References}
%
%
\bibliographystyle{./custom_agsm}
\bibliography{07_references/references.bib}
\end{document}

%% file: 02_nomenclature_and_acronyms/01_nomenclature.tex
\nomenclature{\(\left\lbrace \cdot \right\rbrace\)}{parenthesis for set or operator notation}
\nomenclature{\(\left[ \cdot \right]\)}{parenthesis for matrices, vectors or stochastic moment operations}
\nomenclature{\(\left( \cdot \right)\)}{parenthesis for ordered pair (tuple) or general usage}

%% file: 02_nomenclature_and_acronyms/02_acronyms.tex
\newacronym[
  shortplural={FEMs},
  longplural={finite element methods}
]{fem}{FEM}{finite element method}
\newacronym[
  shortplural={FEs},
  longplural={finite elements}
]{fe}{FE}{finite element}
\newacronym{pe}{PE}{power electronics}
\newacronym[
  shortplural={EMCs},
  longplural={electromagnetic compatibilities}
]{emc}{EMC}{electromagnetic compatibility}
\newacronym[
  shortplural={EMIs},
  longplural={electromagnetic interferences}
]{emi}{EMI}{electromagnetic interference}
\newacronym[
  shortplural={EVs},
  longplural={electric vehicles}
]{ev}{EV}{electric vehicle}
\newacronym[
  shortplural={EUs},
  longplural={European Unions}
]{eu}{EU}{European Union}
\newacronym[
  shortplural={UNECEs},
  longplural={United Nations Economic Commission for Europes}
]{unece}{UNECE}{United Nations Economic Commission for Europe}
\newacronym[
  shortplural={WBKs},
  longplural={Wideband Krigings}
]{wbk}{WBK}{Wideband Kriging}
\newacronym[
  shortplural={GPs},
  longplural={Gaussian processes}
]{gp}{GP}{Gaussian process}
\newacronym[
  shortplural={GPRs},
  longplural={Gaussian process regressions}
]{gpr}{GPR}{Gaussian process regression}
\newacronym[
  shortplural={MOGPs},
  longplural={multi-output Gaussian processes}
]{mogp}{MOGP}{multi-output Gaussian process}
\newacronym[
  shortplural={RBFs},
  longplural={radial basis functions}
]{rbf}{RBF}{radial basis function}
\newacronym[
  shortplural={MOGPRs},
  longplural={multi-output Gaussian process regressions}
]{mogpr}{MOGPR}{multi-output Gaussian process regression}
\newacronym{lsq}{LSQ}{least-squares}
\newacronym{ml}{ML}{maximum likelihood}
\newacronym{map}{MAP}{maximum a posteriori}
\newacronym[
  shortplural={DoEs},
  longplural={design of experiments}
]{doe}{DoE}{design of experiment}
\newacronym{vi}{VI}{variational inference}
\newacronym{gs}{GS}{Gibbs sampling}
\newacronym{mcmc}{MCMC}{Markov Chain Monte Carlo}
\newacronym{hmcmc}{HMCMC}{Hamiltonian Markov Chain Monte Carlo}
\newacronym{iid}{i.i.d.}{independent and identically distributed}
\newacronym{lh}{LH}{latin-hypercube}
\newacronym{maxpro-lhd}{Maxpro-LHD}{maximum projection latin-hypercube design}
\newacronym[
  shortplural={RVs},
  longplural={random variables}
]{rv}{RV}{random variable}
\newacronym[
  shortplural={EESMs},
  longplural={externally excited synchronous motors}
]{eesm}{EESM}{externally excited synchronous motor}
\newacronym[
  shortplural={CADs},
  longplural={computer-aided designs}
]{cad}{CAD}{computer-aided design}
\newacronym[
  shortplural={DUTs},
  longplural={device under tests}
]{dut}{DUT}{device under test}
\newacronym[
  shortplural={DMs},
  longplural={differential modes}
]{dm}{DM}{differential mode}
\newacronym[
  shortplural={CMs},
  longplural={common modes}
]{cm}{CM}{common mode}
\newacronym[
  shortplural={QsoI},
  longplural={quantities of interest}
]{qoi}{QoI}{quantity of interest}
\newacronym{nuts}{NUTS}{No-U-Turn Sampler}
\newacronym{gdm}{GDM}{global difference measure}
\newacronym{fsv}{FSV}{feature selective validation}
\newacronym{loo}{LOO}{leave-one-out}
\newacronym{cv}{CV}{cross-validation}
\newacronym[
  shortplural={ICMs},
  longplural={intrinsic coregionalization models}
]{icm}{ICM}{intrinsic coregionalization model}
\newacronym{iaf}{IAF}{inverse autoregressive flow}
\newacronym{neutra}{NeuTra}{neural transport}
\newacronym{wbg}{WBG}{wide bandgap}
\newacronym{ess}{ESS}{effective sample size}
\newacronym[
  shortplural={ECs},
  longplural={equivalent circuits}
]{ec}{EC}{equivalent circuit}
\newacronym{cmos}{CMOS}{complementary metal-oxide semiconductor}
\newacronym[
  shortplural={MEMSs},
  longplural={micro-electro-mechanical systems}
]{mems}{MEMS}{micro-electro-mechanical system}
\newacronym{fd}{FD}{frequency-domain}
\newacronym{hpc}{HPC}{high-performance computing}
\newacronym[
  shortplural={PCEs},
  longplural={polynomial chaos expansions}
]{pce}{PCE}{polynomial chaos expansion}
\newacronym[
  shortplural={LPMs},
  longplural={lumped parameter models}
]{lpm}{LPM}{lumped parameter model}
\newacronym[
  shortplural={PMSMs},
  longplural={permanent magnet synchronous motors}
]{pmsm}{PMSM}{permanent magnet synchronous motor}
\newacronym[
  shortplural={DAGs},
  longplural={directed acyclic graphs}
]{dag}{DAG}{directed acyclic graph} 
\newacronym[
  shortplural={BNs},
  longplural={Bayesian networks}
]{bn}{BN}{Bayesian network} 
\newacronym[
  shortplural={PDFs},
  longplural={probability density functions}
]{pdf}{PDF}{probability density function} 
\newacronym[
  shortplural={PDs},
  longplural={probability distributions}
]{pd}{PD}{probability distribution} 
\newacronym[
  shortplural={EMPs},
  longplural={effective material parameters}
]{emp}{EMP}{effective material parameter}

%% file: 03_abstract/01_abstract.tex
\textbf{Purpose:} Efficiency-driven faster switching, enabled by wide-bandgap (WBG) semiconductor devices, increases electromagnetic interference (EMI) in electric vehicles (EVs). Mitigating this EMI becomes increasingly complex as the number of system components grows. Therefore, electromagnetic compatibility (EMC) is assessed early in the design phase via system-level simulation of the entire EV powertrain. A key EMC-critical quantity is the impedance of an electric drive machine, forming a major part of the EMI propagation path. Finite element (FE) simulations are typically used but are computationally expensive and limited by insufficient knowledge of material properties.\par
\textbf{Methodology:} We present a surrogate model based Bayesian calibration framework for a 3D full-wave FE simulation of an externally excited synchronous motor (EESM) stator. The effective material parameters (EMPs) are calibrated such that the FE simulation best matches a broadband differential mode (DM) impedance measurement up to $\SI{30}{\mega\hertz}$. A multi-output surrogate model replaces the demanding FE simulation. An orthogonal surrogate model compliant discrepancy term assesses the completeness of the material modeling. Using Hamiltonian Markov Chain Monte Carlo (HMCMC), we construct posterior distributions of the EMPs for three calibration scenarios, which differ in the definition of the EMP ranges.\par
\textbf{Findings:} We found that two of three scenarios achieve good agreement with measurement. However, good DM impedance calibration requires "datasheet-deviating" EMPs, while "datasheet-aligned" ones yield deficient results, indicating that the material models do not fully capture the underlying~physics.\par 
\textbf{Originality:} The proposed calibration methodology enables to estimate unknown material properties from impedance measurements and to evaluate whether chosen material models are sufficient. Although the identified EMPs may differ from datasheet values, they can still be used to estimate the frequency-dependent DM impedance of an EESM stator.\par 
\textbf{Research limitations:} This work is currently limited to calibration of EMPs influencing the EESM stator's DM impedance. Complete machine calibration, including the stator and rotor as well as the common-mode (CM) impedance, is the scope of future research.

%% file: 04_main_content/01_introduction.tex
\section{Introduction}
Since the mid-2010s, automotive industry has been constantly shifting towards sustainable, electric passenger transport, with 25\% of cars sold in 2025 being electric \citep[p. 15]{IEA_GEV_2026}. This trend drives \gls{pe} in \gls{ev} powertrains to be more efficient and power-dense. This is realized through \gls{wbg} semiconductor devices with higher switching frequencies and faster transient slopes \citep{Zhang2020}. However, the rise in switching frequencies and transient slopes increase \gls{emi} which is a significant concern in \gls{ev} powertrains. Mitigating this unwanted \gls{emi} requires understanding its source, propagation path, and sink. While this was traditionally done through measurement-based prototyping, \gls{emc} has been assessed early in the design phase by simulation since the mid-2000s \citep{Lai2006}. This is particularly true at the system level involving entire \gls{ev} powertrains \citep{Spadacini2016,Hillenbrand2017,Gsoels2026,Bohl2026,Gruenwald2026}.\par
A typical \gls{ev} powertrain consists of power electronics, control units, cabling, housings, and electrical drive machines, forming a complex electromagnetic environment \citep{Gsoels2026}. Within this environment, the \gls{emi} propagation path is crucial for the quality of \gls{emi} models in the MHz range and above \citep{Ran1998}. A key \gls{emc}-critical quantity along this propagation path is the impedance of an electric drive machine, which needs to be accurately modeled \citep{Spadacini2016}.\par
\citet{Moreno2021} report various modeling approaches for \gls{emc} analysis of electrical machines in the MHz-range. \Glspl{lpm}, i.e., R-L-C equivalent circuits, are the most common for system-level simulation. Their parameters are obtained from measurements or from \gls{fe} simulations. Measurement-based methods fit the parameters to measured impedance but require a prototype \citep{Moreno2021}. \citet{Wang2010} identifies an \glspl{lpm} for the \gls{cm} and \gls{dm} impedance of an induction motor up to \SI{10}{\mega\hertz}. If no prototype is available, \gls{fe} methods extract the parameters instead. \citet{RuizSarrio2023} identifies an \glspl{lpm} for a multi-conductor transmission line model of a \gls{pmsm}. \gls{fe} methods are an important tool in the pre-prototype design stage \citep{Hewitt2024}. However, they are computationally expensive \citep{Moreno2021}, with computation times in the hours range. They require accurate material properties to describe the impedance, which are, however, often unknown \citep{RuizSarrio2023, Moreno2024, Nayak2026}.\par
Datasheet values of material properties are a reasonable starting point \citep{RuizSarrio2023}, but good agreement between simulation and measurement is difficult to achieve through manual parameter tuning alone. It is only possible if the chosen material modeling approach is adequate for the problem. Therefore, \gls{fe} simulations computing impedances are parametric models in which material parameters, like conductivity $\sigma_{ij}$, permittivity $\varepsilon_{ij}$ and permeability $\mu_{ij}$ in all spatial directions $i,j\in\{x,y,z\}$ are unknown or subject to uncertainty, and the resulting deviations between the simulated and the measured impedance makes model calibration essential.\par
Model calibration adjusts the parameters of a parametric model such that the model best fits an observed \gls{qoi}, such as an impedance measurement. Calibration methods fall into two groups: optimization-based methods that yield parameter point-estimates, such as \gls{lsq}, \gls{ml}, or \gls{map} \citep{Kay1993}, and sampling-based methods that yield the full distribution of the unknown parameters, such as \gls{mcmc} \citep{Brooks2011}. While the former focus on estimating single parameter values, the latter aim for both parameter estimation and an accurate representation of the \gls{qoi}, especially when the model is imperfect \citep{Sung2024}. Consequently, \gls{mcmc} is nowadays increasingly applied in electrical engineering \citep{Neumayer2019,Badeli2021,Fan2026}.\par
\citet{Neumayer2019} infer a spatial permittivity distribution inside a pipe from capacitance measurements, using a state-reduced 2D \gls{fe} forward model. \citet{Badeli2021} infer anatomical parameters of a 3D \gls{fe} human thorax model from impedance measurements, using a \gls{pce} model as a replacement for the electric potential \gls{fe} formulation. \citet{Fan2026} infer R-L-C parameters of high-frequency equivalent circuits of transformer windings from frequency response measurements, using the circuit's analytical transfer function as the forward model.\par
All of these approaches use a surrogate model or model order reduction scheme to reduce the computational cost of the forward simulations. This is because \gls{mcmc}-based parameter inference requires repeated forward model evaluations, typically reaching several thousands \citep[pp.~1,~2]{Paun2021}, which becomes computationally expensive for intensive simulations \citep[pp. 172, 407]{Brooks2011}. However, none of them accounts for model discrepancies arising from a chosen modeling approach. This work presents a Bayesian calibration of the \glspl{emp} ($\sigma_{ij}$, $\varepsilon_{ij}$ and $\mu_{ij}$) of a 3D \gls{fe} simulation of an \gls{eesm} stator. The calibration follows the \citet{Kennedy2001} formulation and is solved via \gls{hmcmc}. A \gls{mogpr} surrogate model replaces the computationally expensive simulation. With this framework we contribute:
\begin{itemize}
  \item The calibration of \glspl{emp} of the 3D \gls{fe} simulation such that the simulated \gls{eesm} stator \gls{dm} impedance matches the measurement up to $\SI{30}{\mega\hertz}$.
  \item An orthogonal Gaussian process discrepancy term that accounts for model inadequacies arising from chosen material modeling. The discrepancy term together with three material calibration scenarios, which differ in the definition of the \gls{emp} ranges, indicate whether the chosen first-order Debye model for the dispersive permeability of the stator core is adequate. The calibrated model predicts the \gls{dm} impedance with good broadband accuracy, with local deviations of 10\,dB around resonances, enabling \gls{emc} simulation of \gls{ev} powertrains at these previously inaccessible frequencies up to $\SI{30}{\mega\hertz}$.
\end{itemize}

%% file: 04_main_content/02_wideband_calibration_problem.tex
\section{Wideband Calibration Problem}
\label{sec:cal_problem}
\begin{figure}[!b]
    \centering
    \begin{minipage}[b]{0.39\textwidth}
        \centering
        \includegraphics[scale=1.0]{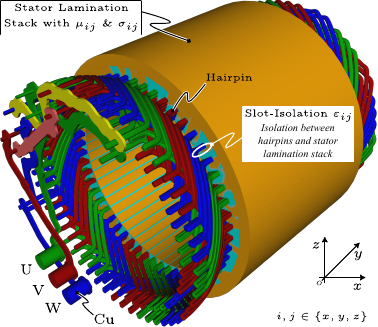}
        \subcaption{}
        \label{fig:3d_cst_stator_geometry}
    \end{minipage}
    \hfill
    \begin{minipage}[b]{0.59\textwidth}
        \centering
        \includegraphics[scale=1.0]{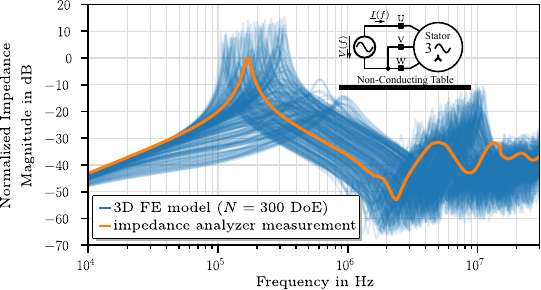}
        \subcaption{}
        \label{fig:surrogate_training_DoE_scaled}
    \end{minipage}
    \caption{Calibration problem for the externally excited synchronous motor (EESM) stator: (a) CAD geometry and (b) \gls{dm} impedance curves from a \emph{CST}-based \citep{DassaultSystemes2026} design of experiment (DoE) (blue) vs. the measurement (orange).}
    \label{fig:wideband_problem}
\end{figure}
The \gls{dut} is the stator of a $\SI{215}{\kilo\watt}$ \gls{eesm} whose \gls{cad} geometry is shown in \figref{fig:3d_cst_stator_geometry}. The \gls{qoi} is its \gls{dm} impedance. The \gls{dm} impedance is measured with an impedance analyzer between phase U and the combined phases V and W up to $\SI{30}{\mega\hertz}$ following the concept outlined in \citet[p.~12]{Moreno2024}. It is compared against a set of \gls{dm} impedance curves produced by the 3D \gls{fe} model of the \gls{eesm} stator \gls{cad} geometry. The curves result from varied \glspl{emp} such as conductivity $\sigma_{ij}$, permittivity $\varepsilon_{ij}$ and permeability $\mu_{ij}$ ($i,j\in\{x,y,z\}$). All \gls{dm} impedance curves are normalized. The measured maximum corresponds to $0\,\text{dB}$. Since the \glspl{emp} are unknown, calibration is needed to find these \glspl{emp} such that the \gls{fe} simulation matches best with measurement. Our calibration is based on the \citet{Kennedy2001} modeling approach, yielding the frequency domain formulation
\begin{align}
\label{eq:freq_dependent_koh_model}
\begin{gathered}
\overbrace{\check{y}_{m}}^{\text{observed}} \coloneq \overbrace{\zeta(f_{m})}^{\text{true}} + \overbrace{\epsilon_{m}}^{\text{noise}} = \overbrace{\mathcal{M}(f_{m},\mathbf{x})}^{\text{simulation model}} + \overbrace{\delta(f_{m})}^{\text{discrepancy}} + \epsilon_{m}
\end{gathered} .
\end{align}
Here, $\check{y}_{m}$ is the measured stator \gls{dm} impedance magnitude at frequency $f_m$ with $m\in\{1,\ldots,M\}$ and $M$ frequency points. The measured quantity $\check{y}_{m}$ is modeled as the true but unknown response $\zeta(\cdot)$ corrupted by additive white Gaussian noise $\epsilon_{m}$. The response $\zeta(\cdot)$ depends on the frequency $f_m$ and is represented by the simulation model $\mathcal{M}(\cdot,\cdot)$, and a systematic discrepancy term $\delta(\cdot)$. The simulation model depends on our unknown \glspl{emp} $\mathbf{x}$ (conductivities, permittivities and permeabilities) and calculates the \gls{dm} impedance, both detailed in \secref{subsec:sim_mdl}. Since the 3D \gls{fe} model $\mathcal{M}(\cdot)$ has simulation times in the hours range, a computationally efficient surrogate model $\widehat{\mathcal{M}}(\cdot)$ is introduced for the simulation model and detailed in \secref{subsec:emu_mdl}. The systematic discrepancy term $\delta(\cdot)$ accounts for differences between $\mathcal{M}(\cdot)$ and the true response $\zeta(\cdot)$, such as simplified model assumptions and is discussed in \secref{subsec:discrepancy_mdl}.
\subsection{Simulation Model}
\label{subsec:sim_mdl}
\figref{fig:3d_cst_stator_geometry} presents the \gls{cad} geometry of the \gls{eesm} stator. The \gls{eesm} stator winding is realized using copper hairpins (green, red, and blue \gls{cad} geometries for each different phase). The stator core (orange) consist of a lamination stack with NO25 non-grain-oriented electrical steel sheets of \(t_{\mathrm{lam}}\coloneq\SI{0.25}{\milli\meter}\) thickness and conductivity of $\SI{1.55e6}{\siemens\per\meter}$ \citep{VoestalpineNO25}. According to \citet{VoestalpineNO25} the relative permeability of NO25 can vary between \mbox{$\mu_{\mathrm{r}} \in [250,  3200]$}. The stator lamination stack is homogenized and modeled as a single body. An effective dispersive material definition is assumed and described by the isotropic complex permeability tensor 
\begin{align}
\label{eq:isotropic_dispersive_mu}
\underline{\boldsymbol{\mu}}(f) \coloneq \mu_0 \; \underline{\mu}_{\mathrm{r}}(f)\;
\diag
\bigl(
1,
1,
1
\bigr),
\end{align}
where $\mu_0$ denotes the free-space permeability, $\underline{\mu}_{\mathrm{r}}(f)$ the dispersive effective relative permeability of the lamination stack and $\diag( \cdot )$ is the operator that maps its arguments to a diagonal matrix. The frequency-dependent relative permeability is modeled using a first-order Debye model \citep{Liu2021}
\begin{align}
\label{eq:first_order_debye_mdl}
\underline{\mu}_{\mathrm{r}}(f) \coloneq \mu_{\mathrm{r},\infty}
+ \frac{\mu_{\mathrm{r},0}-\mu_{\mathrm{r},\infty}}{1+\mathrm{j} f/f_{\mu} }
\end{align}
with the high-frequency relative permeability $\mu_{\mathrm{r},\infty}\approx 1$, the unknown static relative permeability $\mu_{\mathrm{r},0}$, and the unknown relaxation frequency $f_{\mu}$. Equation \eqref{eq:first_order_debye_mdl} can be decomposed into
\begin{align}
\label{eq:complex_permeability}
\underline{\mu}_{\mathrm{r}}(f) \coloneq \mu_{\mathrm{r}}^\prime(f) - \mathrm{j}\; \mu_{\mathrm{r}}^{\prime\prime}(f),
\end{align}
where $\mu_{\mathrm{r}}^\prime(\cdot)$ describes the stored energy and $\mu_{\mathrm{r}}^{\prime\prime}(\cdot)$ the loss component of the complex permeability $\underline{\mu}_{\mathrm{r}}(f)$. According to the isotropy assumption in \eqref{eq:isotropic_dispersive_mu}, direction-dependent permeability, i.e., through different rolling directions, grain orientation and azimuthal orientation of lamination stack along the $y$-direction are not considered. Nevertheless, the aim is to determine whether the simple choice of a first-order Debye model is sufficient to describe the magnetic behavior of the lamination stack.\par
Additionally, the electrical conductivity is modeled as an anisotropic effective material tensor
\begin{align}
\label{eq:conductivity_tensor}
\boldsymbol{\sigma}_{\mathrm{e}} \coloneq
\diag
\bigl(
\sigma_{\mathrm{e},\parallel},
\sigma_{\mathrm{e},\perp},
\sigma_{\mathrm{e},\parallel}
\bigr)
\end{align}
to account for the lamination. Here, $\sigma_{\mathrm{e},\parallel}$ denotes the unknown transversal electric conductivity in the plane of the steel sheets ($xz$-plane) and $\sigma_{\mathrm{e},\perp}$ its respective longitudinal conductivity being orthogonal to the sheet plane ($y$-direction). For simplicity it is assumed, that the conductivities in $x$- and $z$-direction are identical, while the conductivity in longitudinal direction is negligible small, e.g., \mbox{$\sigma_{\mathrm{e},\perp} \coloneq \SI{1}{\micro\siemens\per\meter}$}, due to lamination.\par
The equivalent effect of the coating between hairpins and stator core is modeled by an isotropic permittivity tensor
\begin{align}
\underline{\boldsymbol{\varepsilon}}(f) \coloneq \varepsilon_0 \; \varepsilon_{\mathrm{r}} \; \diag \bigl(1,1,1\bigr),
\end{align}
with unknown relative permittivity $\varepsilon_{\mathrm{r}}$. By combining this information, we define the unknown $D=4$ dimensional \gls{emp} vector
\begin{align}
\label{eq:mat_design_params}
\mathbf{x} \coloneq \left[ \mu_{\mathrm{r},0}, f_{\mu}, \varepsilon_{\mathrm{r}}, \sigma_{\mathrm{e},\parallel} \right]^\top .
\end{align}
Thus, the \gls{dm} \gls{eesm} stator impedance is dependent on the unknown \glspl{emp} $\mathbf{x}$ and is defined by the forward simulation model
\begin{align}
\label{eq:sim_mdl}
\boldsymbol{\mathcal{M}}:\mathbb{R}^{D}\rightarrow\mathbb{R}^{M},
\qquad
\mathbf{y}
\coloneq
\boldsymbol{\mathcal{M}}(\mathbf{x})
=
\begin{bmatrix}
\mathcal{M}(f_1,\mathbf{x}) & \ldots &  \mathcal{M}(f_M,\mathbf{x})\\
\end{bmatrix}^\top
\end{align}
which returns the $\mathrm{log}_{10}$ of the \gls{dm} impedance. The forward simulation is realized by the commercial \gls{fe} tool \emph{CST Studio Suite} \citep{DassaultSystemes2026} and its \gls{fd} solver. Here, $M$ is the number of logarithmically equidistant frequency points at which $\boldsymbol{\mathcal{M}}(\mathbf{x})$ is evaluated. Solving the meshed forward problem (1,697,917 mesh cells) requires on average $\SI{48}{\minute}$ on an \gls{hpc} system equipped with an AMD EPYC 7H12 64-Core Processor and $\SI{256}{\giga\byte}$ of RAM.
\subsection{Wideband Surrogate Model}
\label{subsec:emu_mdl}
Model calibration requires many simulations, which are too costly at $\SI{48}{\minute}$ of evaluation time of the \gls{fe} model \eqref{eq:sim_mdl}. Therefore, we construct a surrogate model using a \gls{wbk} \citep{Rezende2024} formulation. The numerically efficient surrogate model is built from training data generated by a previously conducted \gls{doe} (\secref{subsubsec:os_doe}). Using the training data and the \gls{wbk} formulation in \secref{subsubsec:surrogate_formulation}, we train the model and evaluate its accuracy in \secref{subsubsec:surrogate_accuracy}.
\subsubsection{One-Shot Design of Experiment}
\label{subsubsec:os_doe}
We construct the surrogate model using a space-filling \gls{maxpro-lhd} to explore the design space with $N$ \gls{doe} samples of the \glspl{emp}. The resulting \gls{maxpro-lhd} is given as
\begin{align}
  \mathbf{X} =
  \begin{bmatrix}
      \mathbf{x}^{(1)}
      &
      \cdots
      &
      \mathbf{x}^{(N)}
  \end{bmatrix}
  \sim \text{Maxpro-LHD}(D,N),
\end{align}
where $D$ is the number of \glspl{emp}. The \gls{doe} is implement using the Python package \emph{PyDOE} \citep{Lee2017}.\par
The simulation model \eqref{eq:sim_mdl} is evaluated for all $\mathbf{x}^{(i)}\in \mathbb{R}^D$ \gls{emp} samples, with $i\in\{1,\ldots,N\}$, resulting in
\begin{align}
  \mathbf{y}^{(i)}
  \coloneq
  \begin{bmatrix}
      \mathcal{M}(f_1,\mathbf{x}^{(i)})
      &
      \cdots
      &
      \mathcal{M}(f_M,\mathbf{x}^{(i)})
  \end{bmatrix}^\top,
\end{align}
where all $M=351$-dimensional normalized \gls{dm} impedance observations $\mathbf{y}^{(i)}$ are collected column-wise in
\begin{align}
  \mathbf{Y}
  \coloneq
  \begin{bmatrix}
      \mathbf{y}^{(1)}
      &
      \cdots
      &
      \mathbf{y}^{(N)}
  \end{bmatrix}
  \in\mathbb{R}^{M\times N}.
\end{align}
The $N=300$-sample \gls{doe} is performed for the $D=4$ \glspl{emp} defined in \eqref{eq:mat_design_params}, using the wide ranges of parameters listed in \tabref{tab:design_parameter_ranges}. Executed sequentially, it requires approximately 10 days on the \gls{hpc} system described in \secref{subsec:sim_mdl}. The parameter ranges in \tabref{tab:design_parameter_ranges} produce \gls{dm} impedance curves with broadband variations from $\SI{10}{\kilo\hertz}$ to $\SI{30}{\mega\hertz}$, as visible in \figref{fig:surrogate_training_DoE_scaled}. Thus, the measured impedance falls within the \gls{fe} model's solution manifold, suggesting that suitable parameter combinations might exist to match simulation with measurement. This is evaluated in the calibration concept and results of \secref{sec:bayesian_inf_formulation} and \secref{sec:bayesian_cal_results}.
\begin{table}[!h]
  \centering
  \caption{Surrogate model parameter ranges. $x_{d}^{\mathrm{l}}$ denotes the $d$-th lower and $x_{d}^{\mathrm{u}}$ the $d$-th upper bound.}
  \vspace{4pt}
  \label{tab:design_parameter_ranges}
  {\scriptsize
  \begin{tabular}{
      l|
      c|
      c|
      c
      c
    }
    \toprule
     $d$ & Param. & Unit & $x_{d}^{\mathrm{l}}$ & $x_{d}^{\mathrm{u}}$ \\
    \midrule
    1 & $\mu_{\mathrm{r,0}}$ & - & \num{1e2} & \num{4e3} \\
    2 & $f_{\mu}$ & $\si{\hertz}$ & \num{8e2} & \num{1.25e6} \\
    3 & $\varepsilon_{\mathrm{r}}$ & - & \num{0.75} & \num{12} \\
    4 & $\sigma_{\mathrm{e},\parallel}$ & $\si{\siemens/\meter}$ & \num{1e2} & \num{1e7} \\
    \bottomrule
  \end{tabular}
  }
\end{table}
\subsubsection{Surrogate Model Formulation}
\label{subsubsec:surrogate_formulation}
We use the surrogate model
\begin{align}
\label{eq:emu_mdl}
\widehat{\boldsymbol{\mathcal{M}}}:\mathbb{R}^{D}\rightarrow\mathbb{R}^{M},
\qquad
\hat{\mathbf{y}}
\coloneq
\widehat{\boldsymbol{\mathcal{M}}}(\mathbf{x})
=
\begin{bmatrix}
\widehat{\mathcal{M}}(f_1,\mathbf{x}) & \ldots &  \widehat{\mathcal{M}}(f_M,\mathbf{x})\\
\end{bmatrix}^\top
\end{align}
to approximate the \gls{dm} stator impedance simulation model $\boldsymbol{\mathcal{M}}(\mathbf{x})$. It is implemented using a noise-free \gls{wbk} \citep{Rezende2024} surrogate model (a.k.a. \gls{mogpr} model). This predictive model is described by the $M$-variate distribution
\begin{align}
  \label{eq:wbk_predictive_dist}
  \mathscr{p}(\hat{\mathbf{y}} \mid \mathbf{X}, \mathbf{Y}, \mathbf{x}) \coloneq \mathcal{N}\big(\hat{\mathbf{y}} \mid \boldsymbol{\mu}_{\hat{\mathbf{y}}}(\boldsymbol{\nu}_{\mathbf{x}}(\mathbf{x})), \boldsymbol{\Sigma}_{\hat{\mathbf{y}}}(\boldsymbol{\nu}_{\mathbf{x}}(\mathbf{x})) \big),
\end{align}
where $\hat{\mathbf{y}}$ is the \gls{rv} describing the \gls{dm} stator impedance. The functions $\boldsymbol{\mu}_{\hat{\mathbf{y}}}(\cdot)$ and $\boldsymbol{\Sigma}_{\hat{\mathbf{y}}}(\cdot)$ denote the corresponding \gls{emp}-dependent predictive mean and covariance, respectively.\par
The function $\boldsymbol{\nu}_{\mathbf{x}}(\cdot)$ performs feature scaling of the $D$-dimensional \gls{emp} vector $\mathbf{x}$ to improve numerical stability. It first maps $\mathbf{x}$ to $\mathrm{log}_{10}$-space and subsequently to a $D$-dimensional unit hypercube. The resulting feature-scaled \gls{emp} vector is denoted by
\begin{align}
  \label{eq:parameter_feature_scaling}
  \tilde{\mathbf{x}} \coloneq \boldsymbol{\nu}_{\mathbf{x}}(\mathbf{x}) \in [0,1]^D.
\end{align}
The matrix $\widetilde{\mathbf{X}}=\left[\tilde{\mathbf{x}}^{(1)},\ldots,\tilde{\mathbf{x}}^{(N)}\right]\in\mathbb{R}^{D\times N}$ contains the feature-scaled \gls{maxpro-lhd} samples, while the corresponding simulation outputs are collected in $\mathbf{Y}\in\mathbb{R}^{M\times N}$. The \gls{wbk} surrogate model is trained using these input-output data.\par
For the \gls{wbk} surrogate model, the predictive mean and covariance are given by
\begin{align}
  \boldsymbol{\mu}_{\hat{\mathbf{y}}}(\tilde{\mathbf{x}}) & \coloneq \mathbf{m}_{\mathbf{Y}}(\hat{\boldsymbol{\theta}}) + \left( \mathbf{Y} - \mathbf{m}_{\mathbf{Y}}(\hat{\boldsymbol{\theta}})\, \boldsymbol{1}^\top \right) \left[ k_x(\widetilde{\mathbf{X}},\widetilde{\mathbf{X}};\hat{\boldsymbol{\theta}}) \right]^{-1}\, k_x(\widetilde{\mathbf{X}},\tilde{\mathbf{x}};\hat{\boldsymbol{\theta}})\in \mathbb{R}^{M}, \label{eq:predictive_mu} \\
  \boldsymbol{\Sigma}_{\hat{\mathbf{y}}}(\tilde{\mathbf{x}}) & \coloneq \boldsymbol{\Sigma}_{\mathbf{Y}}(\hat{\boldsymbol{\theta}}) \cdot \left( k_x(\tilde{\mathbf{x}},\tilde{\mathbf{x}};\hat{\boldsymbol{\theta}}) - k_x(\widetilde{\mathbf{X}},\tilde{\mathbf{x}};\hat{\boldsymbol{\theta}})^\top \, \left[ k_x(\widetilde{\mathbf{X}},\widetilde{\mathbf{X}};\hat{\boldsymbol{\theta}}) \right]^{-1}\, k_x(\widetilde{\mathbf{X}},\tilde{\mathbf{x}};\hat{\boldsymbol{\theta}}) \right)\in \mathbb{R}^{M\times M}. \label{eq:predictive_cov}
\end{align}
Here, $\boldsymbol{1}\coloneq[1,\ldots,1]^\top$ denotes an $N$-dimensional vector of ones. The function $k_x(\cdot,\cdot;\cdot)$ defines the Kriging kernel of the \gls{wbk} surrogate model. The input-independent mean vector $\mathbf{m}_{\mathbf{Y}}(\cdot)$, the output covariance $\boldsymbol{\Sigma}_{\mathbf{Y}}(\cdot)$, and the kernel function $k_x(\cdot,\cdot;\cdot)$ are parameterized by the trained hyperparameters $\hat{\boldsymbol{\theta}}$.\par
The hyperparameters $\hat{\boldsymbol{\theta}}$ are obtained by maximizing the marginal log-likelihood
\begin{align}
  \label{eq:log_marginal_likelihood}
  \mathscr{l} \left(\boldsymbol{\theta} \right) \coloneq & -\frac{M}{2} \log \left( \det \left( k_x(\widetilde{\mathbf{X}},\widetilde{\mathbf{X}};\boldsymbol{\theta}) \right) \right) -\frac{N M}{2} \log \left( 2\pi \right) -\frac{N}{2} \log \left( \det \left( \boldsymbol{\Sigma}_{\mathbf{Y}}(\boldsymbol{\theta}) \right) \right) \nonumber \\
  & -\frac{1}{2} \tr \left( \left[ \boldsymbol{\Sigma}_{\mathbf{Y}}(\boldsymbol{\theta})\right]^{-1} \left( \mathbf{Y} - \mathbf{m}_{\mathbf{Y}}(\boldsymbol{\theta})\, \boldsymbol{1}^\top \right) \left[ k_x(\widetilde{\mathbf{X}},\widetilde{\mathbf{X}};\boldsymbol{\theta}) \right]^{-1} \left( \mathbf{Y} - \mathbf{m}_{\mathbf{Y}}(\boldsymbol{\theta})\, \boldsymbol{1}^\top \right)^\top \right)
\end{align}
of a matrix-variate normal distribution \citep[pp. 55-56]{Gupta1999} where the corresponding \gls{ml} estimates are
\begin{align}
  \mathbf{m}_{\mathbf{Y}}\left(\boldsymbol{\theta} \right) & \coloneq \frac{\mathbf{Y}\, \left[ k_x(\widetilde{\mathbf{X}},\widetilde{\mathbf{X}};\boldsymbol{\theta}) \right]^{-1}\, \mathbf{1}}{\mathbf{1}^\top\, \left[ k_x(\widetilde{\mathbf{X}},\widetilde{\mathbf{X}};\boldsymbol{\theta}) \right]^{-1}\, \mathbf{1}} \\
  \boldsymbol{\Sigma}_{\mathbf{Y}}\left(\boldsymbol{\theta} \right) & \coloneq \diag\left( \frac{\left( \mathbf{Y} - \mathbf{m}_{\mathbf{Y}}(\boldsymbol{\theta})\, \boldsymbol{1}^\top \right)\, \left[ k_x(\widetilde{\mathbf{X}},\widetilde{\mathbf{X}};\boldsymbol{\theta}) \right]^{-1}\, \left( \mathbf{Y} - \mathbf{m}_{\mathbf{Y}}(\boldsymbol{\theta})\, \boldsymbol{1}^\top \right)^\top}{N}\right).
\end{align}
The maximization is performed using a \gls{ml} estimation routine. Further details of this particular surrogate model can be found in \citet{Rezende2024} and \citet{Gsoels2026}.
\subsubsection{Surrogate Model Accuracy}
\label{subsubsec:surrogate_accuracy}
\begin{figure}[!h]
  \centering
  \includegraphics[scale=1.0]{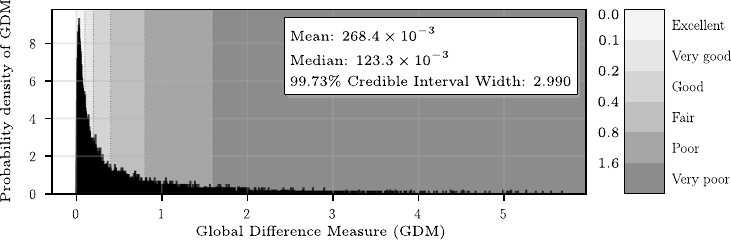}
  \caption{Probability density function (PDF) of all global difference measure (GDM) values evaluated for $N$ leave-one-out cross-validation samples.}
  \label{fig:GDM_epdf_of_loocv}
\end{figure}

The surrogate model accuracy is assessed for the $N\coloneq 300$ \gls{doe} samples from \secref{subsubsec:os_doe}. The accuracy is quantified by the \gls{gdm}
\begin{align}
  \mathbf{GDM} \coloneq \mathrm{f}(\mathbf{y},\boldsymbol{\mu}_{\hat{\mathbf{y}}}) \in \mathbb{R}_{\ge 0}^{M}
\end{align}
of the standardized \gls{fsv} tool~\citep[pp.~24-31]{IEEEElectromagnetic2022}. This metric compares the simulation model response $\mathbf{y}$ against the mean surrogate model response $\boldsymbol{\mu}_{\hat{\mathbf{y}}}$ and is computed using \gls{loo} \gls{cv} \citep[p. 111]{Rasmussen2006} for all $N$ training samples. Although \gls{loo}-\gls{cv} is computationally demanding due to model retraining, it can be efficiently implemented for our \gls{wbk} model using a Schur-complement based partitioning. The conditional covariance for prediction is then calculated without retraining the model. This approach is demonstrated for scalar \gls{gpr} models in \citet[pp. 116-118]{Rasmussen2006} and for \gls{mogpr} models in \citet[p. 162]{Truffinet2024}. The latter is applied to our used \gls{wbk} surrogate model. The \gls{gdm} accuracy result is shown in \figref{fig:GDM_epdf_of_loocv}. According to \gls{fsv}'s qualitative accuracy definition \citep[p.~29]{IEEEElectromagnetic2022}, the mean accuracy of "good" is achieved, as it falls in \gls{fsv}'s accuracy range of $[0.2,0.4]$. This model accuracy is sufficient for the calibration presented later.
\subsection{Orthogonal Wideband Surrogate Model Compliant Discrepancy Model}
\label{subsec:discrepancy_mdl}
The surrogate model \eqref{eq:emu_mdl} replaces the simulation model \eqref{eq:sim_mdl} inside the observation model \eqref{eq:freq_dependent_koh_model} and appears additively with the discrepancy term $\delta(\cdot)$. This additive structure allows the model and the discrepancy term to cancel each other out, leading to parameter identifiability problems \citep{Plumlee2018}. To decouple them, we impose an orthogonality condition such that the discrepancy is not part of the surrogate model's manifold.\par
The following derivation of an orthogonal surrogate model compliant discrepancy term is based on the orthogonal discrepancy concept by \citet{Plumlee2017} and \citet{Plumlee2018}. The starting point is the expected value of the vector-valued \citet{Kennedy2001} observation model from \eqref{eq:freq_dependent_koh_model}, which is described by
\begin{align}
  \label{eq:emu_koh_expected_value}
  \expect[\check{\mathbf{y}}] = \expect[\widehat{\boldsymbol{\mathcal{M}}}(\mathbf{x}) + \boldsymbol{\delta} + \boldsymbol{\epsilon}],\quad \boldsymbol{\epsilon}\sim\mathcal{N}(\mathbf{0},\boldsymbol{\Sigma}_{\boldsymbol{\epsilon}}).
\end{align}
The discrepancy $\boldsymbol{\delta}\in\mathbb{R}^M$ can be described by decomposing it into tangential $\boldsymbol{\delta}_{\parallel}$ and normal $\boldsymbol{\delta}_{\perp}$ components using the projection identity 
\begin{align}
  \label{eq:projection_identity}
  \mathbf{I}\,\boldsymbol{\delta} \equiv \left( \mathbf{P}_{\parallel} + \mathbf{P}_{\perp} \right)\boldsymbol{\delta} \quad\Rightarrow\quad \boldsymbol{\delta} = \boldsymbol{\delta}_{\parallel} + \boldsymbol{\delta}_{\perp}
\end{align}
from \citet[Sec.~2.5.1, p.~82]{Golub2013}. We assume that the tangential component is defined by a affine mapping 
\begin{align}
  \label{eq:tangential_delta_component}
  \boldsymbol{\delta}_{\parallel} \coloneqq \mathbf{J}_{\boldsymbol{\mu}} \Delta\mathbf{x},
\end{align}
where $\Delta\mathbf{x}\in \mathbb{R}^{D}$ describes the change of the unknown \gls{emp} vector and $\mathbf{J}_{\boldsymbol{\mu}}$ the Jacobian matrix 
\begin{align}
\mathbf{J}_{\boldsymbol{\mu}}(\mathbf{x}) = \left[ \mathbf{j}_1,\ldots,\mathbf{j}_D \right] \coloneqq \left. \frac{\partial \boldsymbol{\mu}_{\hat{\mathbf{y}}}(\boldsymbol{\nu}_{\mathbf{x}}(\mathbf{x}'))}{\partial \mathbf{x}'}\right|_{\mathbf{x}} \in\mathbb{R}^{M\times D}
\end{align}
of $\boldsymbol{\mu}_{\hat{\mathbf{y}}}(\boldsymbol{\nu}_{\mathbf{x}}(\mathbf{x}))$ evaluated for $\mathbf{x}$. The computation of the Jacobian matrix can be efficiently implemented using automatic differentiation. Consequently, \eqref{eq:emu_koh_expected_value} is rewritten as
\begin{align}
  \expect[\check{\mathbf{y}}] = \boldsymbol{\mu}_{\hat{\mathbf{y}}}(\boldsymbol{\nu}_{\mathbf{x}}(\mathbf{x})) + \mathbf{J}_{\boldsymbol{\mu}} \Delta\mathbf{x} + \boldsymbol{\delta}_{\perp}
\end{align}
by using \eqref{eq:projection_identity} and \eqref{eq:tangential_delta_component}. The first two terms $\boldsymbol{\mu}_{\hat{\mathbf{y}}}(\boldsymbol{\nu}_{\mathbf{x}}(\mathbf{x}))$ and $\mathbf{J}_{\boldsymbol{\mu}} \Delta\mathbf{x}$ are consistent with a Taylor expansion
\begin{align}
  \boldsymbol{\mu}_{\hat{\mathbf{y}}}(\boldsymbol{\nu}_{\mathbf{x}}(\mathbf{x} + \Delta\mathbf{x})) \approx \boldsymbol{\mu}_{\hat{\mathbf{y}}}(\boldsymbol{\nu}_{\mathbf{x}}(\mathbf{x})) + \mathbf{J}_{\boldsymbol{\mu}} \Delta\mathbf{x}
\end{align}
of the surrogate model's predictive mean in the expansion point $\mathbf{x}$, which is truncated after the linear term. Thus, the surrogate model output change (tangential component of the discrepancy term $\boldsymbol{\delta}_{\parallel}$) is approximately described by the \gls{emp} change $\Delta\mathbf{x}$ and the Jacobian matrix $\mathbf{J}_{\boldsymbol{\mu}}(\mathbf{x})$. For changes in $\Delta\mathbf{x}$ to contribute exclusively to the tangential component of $\boldsymbol{\delta}$, the condition
\begin{align}
\label{eq:delta_ortho_condition}
\mathbf{J}_{\boldsymbol{\mu}}^\top \boldsymbol{\delta}_{\perp} = \mathbf{0}
\end{align}
must hold. In other words, the discrepancy term is intended to take effect only there, where the model does not contribute. Hence, we are seeking for the projection matrix $\mathbf{P}_{\perp}$ which induces $\boldsymbol{\delta}_{\perp} = \mathbf{P}_{\perp}\, \boldsymbol{\delta}$.\par
Next, we assume an arbitrary initial discrepancy term, which is described by a \gls{gp} prior
\begin{align}
  \delta(f) \sim \mathcal{GP} \left(0, k_\delta(\nu_{f}(f),\nu_{f}(f');\sigma_\delta^2,\ell_\delta) \right),
\end{align}
where $\nu_{f}(\cdot)$ is the function performing feature scaling on the frequency $f$ (similar to \eqref{eq:parameter_feature_scaling}). We use an \gls{rbf} kernel for the discrepancy \gls{gp}
\begin{align}
  \label{eq:delta_kernel_fun}
  k_\delta(\tilde{f},\tilde{f}';\sigma_\delta^2,\ell_\delta)\coloneq \sigma_\delta^2\, \exp\left( - \frac{| \tilde{f} - \tilde{f}' |^2}{2 \ell_\delta^2} \right),
\end{align}
with the length scale $\ell_{\delta}$ and the signal variance $\sigma_{\delta}^2$ summarized as kernel hyperparameters $\boldsymbol{\theta}_{\delta}\coloneq [\sigma_{\delta}^2 , \ell_{\delta}]^\top$. The kernel defines the \gls{gp}'s prior distribution
\begin{align}
\label{eq:original_delta_distribution}
\boldsymbol{\delta} \mid \mathbf{f}, \sigma_{\delta}^2, \ell_{\delta} \sim \mathcal{N} \left( \mathbf{0}, \mathbf{K}_{\boldsymbol{\delta}} \right),
\end{align}
where the covariance $\mathbf{K}_{\boldsymbol{\delta}}$ is positive definite ($\mathbf{K}_{\boldsymbol{\delta}}\succ 0$) and $\mathbf{f}\coloneq [f_1,\ldots,f_M]^\top$ is the frequency vector. However, since $\boldsymbol{\delta}$ consists of tangential and normal components and is not yet related to the orthogonality condition in \eqref{eq:delta_ortho_condition}, we want to find the discrepancy term that satisfies the orthogonality condition, which can be expressed in the \gls{lsq} sense through the optimization problem 
\begin{align}
\begin{aligned}
\min_{\boldsymbol{\delta}_{\perp}} \;\; & \left( \boldsymbol{\delta}_{\perp} - \boldsymbol{\delta} \right)^\top \left( \boldsymbol{\delta}_{\perp} - \boldsymbol{\delta} \right) & \text{} \\
\text{s.t.} \;\;  & \mathbf{J}_{\boldsymbol{\mu}}^\top \boldsymbol{\delta}_{\perp} = \mathbf{0} & \, \text{}
\end{aligned}.
\end{align}
This constraint optimization problem can be rewritten in an unconstrained one by Lagrange multipliers, which results in
\begin{align}
\mathscr{L}(\boldsymbol{\delta}_{\perp},\boldsymbol{\lambda}) = \left( \boldsymbol{\delta}_{\perp} - \boldsymbol{\delta} \right)^\top \left( \boldsymbol{\delta}_{\perp} - \boldsymbol{\delta} \right) + 2 \boldsymbol{\lambda}^\top \mathbf{J}_{\boldsymbol{\mu}}^\top \boldsymbol{\delta}_{\perp}.
\end{align}
Solving the extremum problem $\nabla_{\boldsymbol{\delta}_{\perp},\boldsymbol{\lambda}} \mathscr{L} \overset{!}{=} \mathbf{0}$ results in the orthogonal discrepancy
\begin{align}
\boldsymbol{\delta}_{\perp} = \left[ \mathbf{I} - \mathbf{J}_{\boldsymbol{\mu}} \left( \mathbf{J}_{\boldsymbol{\mu}}^\top \mathbf{J}_{\boldsymbol{\mu}} \right)^{-1} \mathbf{J}_{\boldsymbol{\mu}}^\top \right] \boldsymbol{\delta} .
\end{align}
The part in square-brackets corresponds to the projection operator
\begin{align}
\label{eq:projection_matrix}
\mathbf{P}_{\perp} = \mathbf{I} - \mathbf{P}_{\parallel} = \mathbf{I} - \mathbf{J}_{\boldsymbol{\mu}} \left( \mathbf{J}_{\boldsymbol{\mu}}^\top \mathbf{J}_{\boldsymbol{\mu}} \right)^{-1} \mathbf{J}_{\boldsymbol{\mu}}^\top \in\mathbb{R}^{M\times M}
\end{align}
which generates the normal component $\boldsymbol{\delta}_{\perp}$ of the discrepancy term $\boldsymbol{\delta}$.\par
By using the affine transformation $\boldsymbol{\delta}_{\perp} = \mathbf{P}_{\perp} \boldsymbol{\delta}$, the covariance from \eqref{eq:original_delta_distribution} changes to
\begin{align}
  \label{eq:ortho_delta_cov}
  \mathbf{K}_{\boldsymbol{\delta},\perp}(\mathbf{x},\sigma_{\delta}^2, \ell_{\delta}) \coloneq \mathbf{P}_{\perp}(\mathbf{x})\, \mathbf{K}_{\boldsymbol{\delta}}(\sigma_{\delta}^2, \ell_{\delta})\, \mathbf{P}_{\perp}^\top(\mathbf{x}) + \sigma_{\perp}\, \mathbf{I} .
\end{align}
The additional term $\sigma_{\perp}\, \mathbf{I}$, regularizes the covariance. It is used to guarantee $(\mathbf{P}_{\perp} \mathbf{K}_{\boldsymbol{\delta}} \mathbf{P}_{\perp}^\top + \sigma_{\perp}\, \mathbf{I}) \succ 0$. This is necessary because transforming a positive-definite matrix $\mathbf{K}_{\boldsymbol{\delta}}$ by an arbitrary matrix yields a positive-semidefinite matrix whose rank equals that of the transformation matrix \citep[Observation~7.1.8, p.~431]{Horn2013}. The projection matrix $\mathbf{P}_{\perp}$ produces a singular positive-semidefinite covariance matrix $\mathbf{P}_{\perp} \mathbf{K}_{\boldsymbol{\delta}} \mathbf{P}_{\perp}^\top$. This is because $\mathbf{P}_{\perp}$ is rank-deficient according to \eqref{eq:projection_matrix}, i.e., $\rank(\mathbf{P}_{\perp})<M$, since $\rank(\mathbf{P}_{\parallel})>0$ as $\mathbf{P}_{\parallel}$ is not the zero matrix. This yields the orthogonal discrepancy \gls{gp}
\begin{align}
\label{eq:ortho_delta}
\boldsymbol{\delta}_{\perp} \mid \mathbf{f}, \mathbf{x}, \sigma_{\delta}^2, \ell_{\delta} \sim \mathcal{N} \left( \mathbf{0}, \mathbf{K}_{\boldsymbol{\delta},\perp}(\mathbf{x},\sigma_{\delta}^2, \ell_{\delta}) \right) .
\end{align}

%% file: 04_main_content/03_bayesian_inference_formulation.tex
\section{Bayesian Inference for Model Calibration}
\label{sec:bayesian_inf_formulation}
We use a vector-valued formulation of the \citet{Kennedy2001} calibration model, the surrogate model from~\eqref{eq:emu_mdl}, and the orthogonal surrogate model compliant discrepancy term from \eqref{eq:ortho_delta}, which yields the calibration model
\begin{align}
\label{eq:vector_valued_koh}
\check{\mathbf{y}} = \widehat{\boldsymbol{\mathcal{M}}}(\mathbf{x}) + \boldsymbol{\delta}_{\perp} + \boldsymbol{\epsilon},
\end{align}
where both the measured quantity $\check{\mathbf{y}}$ and the surrogate model output $\widehat{\boldsymbol{\mathcal{M}}}(\mathbf{x})$ are quantities of the $\mathrm{log}_{10}$-space.  
The right-hand side of \eqref{eq:vector_valued_koh} is described exclusively by \glspl{rv}, which yields the hierarchical statistical formulation
\begin{align}
 \label{eq:hierarchical_stat_formulation}
 \begin{aligned}
 \check{\mathbf{y}} \mid \hat{\mathbf{y}},  \boldsymbol{\delta}_{\perp}, \sigma_{\epsilon}
 &\sim \mathcal{N}(\hat{\mathbf{y}} + \boldsymbol{\delta}_{\perp}, \sigma_{\epsilon}^2\; \mathbf{I}) & \text{(data likelihood)} \\
 \hat{\mathbf{y}} \mid \mathbf{X}, \mathbf{Y}, \mathbf{x} & \sim \mathcal{N}\big( \boldsymbol{\mu}_{\hat{\mathbf{y}}}(\boldsymbol{\nu}_{\mathbf{x}}(\mathbf{x})), \boldsymbol{\Sigma}_{\hat{\mathbf{y}}}(\boldsymbol{\nu}_{\mathbf{x}}(\mathbf{x})) \big) & \text{(surrogate model predictive distribution)} \\
 \boldsymbol{\delta}_{\perp} \mid \mathbf{f}, \mathbf{x}, \sigma_{\delta}^2, \ell_{\delta} & \sim \mathcal{N}\big( \mathbf{0}, \mathbf{K}_{\boldsymbol{\delta},\perp}(\mathbf{x},\sigma_{\delta}^2, \ell_{\delta}) \big) & \text{(orthogonal $\delta$-GP prior)} \\
 x_{d} & \sim \mathrm{LogUniform}\big( x_{d}^{\mathrm{l}},x_{d}^{\mathrm{u}} \big) \quad\forall d & \text{(design parameter priors)} \\
  \sigma_{\delta}^2 & \sim \mathrm{LogUniform}\big( 10^{-7},10^{7} \big) & \text{(variance prior of $\delta$-kernel)} \\
 \ell_{\delta} & \sim \mathrm{LogUniform}\big( 10^{-7},10^{7} \big)  & \text{(length-scale prior of $\delta$-kernel)} \\
 \sigma_{\epsilon} & \sim \mathrm{Beta}\big( 1.2, 20.8 \big) & \text{(noise prior)}
 \end{aligned}
\end{align}
for the calibration problem. The statistical formulation is hierarchical because the $M$-dimensional Gaussian data likelihood parameters are themselves treated as \glspl{rv} with their own distributions.\par
The mean of the data likelihood is the sum of the surrogate model's random response $\hat{\mathbf{y}}$ and the orthogonal discrepancy $\boldsymbol{\delta}_{\perp}$. The response $\hat{\mathbf{y}}$ follows an $M$-dimensional Gaussian (posterior) predictive distribution of the \gls{wbk} model~\eqref{eq:wbk_predictive_dist}. The moments $\boldsymbol{\mu}_{\hat{\mathbf{y}}}(\cdot)$ \eqref{eq:predictive_mu} and $\boldsymbol{\Sigma}_{\hat{\mathbf{y}}}(\cdot)$ \eqref{eq:predictive_cov} are hierarchically dependent on the unknown \glspl{emp} $x_d$, for which LogUniform priors are defined, since we have no prior knowledge of their distribution. The lower $x_{d}^{\mathrm{l}}$ and upper $x_{d}^{\mathrm{u}}$ bounds of the \glspl{emp}' LogUniform priors are defined in \secref{sec:bayesian_cal_results}. The covariance $\mathbf{K}_{\boldsymbol{\delta},\perp}(\cdot)$ of the $M$-dimensional orthogonal discrepancy \gls{gp} prior is hierarchically dependent on the unknown \glspl{emp} $x_d$ and the discrepancy kernel's hyperparameters $\sigma_{\delta}^2$ and $\ell_{\delta}$. For the hyperparameters $\sigma_{\delta}^2$ and $\ell_{\delta}$, we assume LogUniform priors with wide ranges since we also have no prior knowledge of their distribution.\par 
The data likelihood's covariance is described by additive white Gaussian noise $\sigma_{\epsilon}^2\,\mathbf{I}$, i.e., with \gls{iid} noise components. The noise standard deviation $\sigma_{\epsilon}$ is modeled by a Beta prior. The Beta prior is defined by its $\alpha\coloneq 1.2$ and $\beta\coloneq 20.8$ parameters, placing the distribution mode at $\nicefrac{(\alpha - 1)}{(\alpha + \beta - 2)}=0.01$. Thereby we provide information that the noise is most probably low with values around $0.01$ in the $\mathrm{log}_{10}$-space of the impedance quantity.  This $[0,1]$-supported distribution is suitable because the \gls{dm} stator impedance (\figref{fig:surrogate_training_DoE_scaled}) is virtually noise-free and $\sigma_{\epsilon}= 0.01$ falls within its support. Although the problem is noise-free, the noise standard deviation $\sigma_{\epsilon}$ serves an additional function as regularizer, which is beneficial for the definiteness of the covariance matrix which becomes apparent later in \eqref{eq:res_marginalization_int}.\par
The hierarchical statistical model in \eqref{eq:hierarchical_stat_formulation} defines our probabilistic inverse problem for our calibration task, where we seek the \glspl{emp} $\mathbf{x}$ that best explain the measured \gls{eesm} stator impedance $\check{\mathbf{y}}$. As a first step, we represent \eqref{eq:hierarchical_stat_formulation} as a \gls{dag} \citep[p. 37]{Koller2009}. The \gls{dag} provides an efficient description of the probabilistic inverse problem, which is described by a \gls{bn} \citep{BenGal2008}. \Glspl{bn} based on \glspl{dag} provide a graphical description of a domain with uncertainty \citep{BenGal2008}.
\begin{figure}[!h]
  \centering
  \includegraphics[scale=1.0]{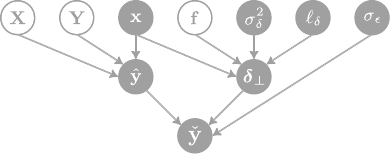}
  \caption{Directed acyclic graph (DAG) of the hierarchical statistical problem in \eqref{eq:hierarchical_stat_formulation}.}
  \label{fig:bnet_of_stat_form}
\end{figure}
\figref{fig:bnet_of_stat_form} shows this graphically, where the nodes correspond to all \glspl{rv} in \eqref{eq:hierarchical_stat_formulation}, and its edges encode the probabilistic dependencies between them \citep{BenGal2008}. According to \citet[p.~62]{Koller2009}, the \gls{dag}-based \gls{bn} factorizes the joint distribution of all \glspl{rv} as
\begin{align}
  \label{eq:bn_chain_rule}
  \mathscr{p}( \alpha_1,\ldots,\alpha_{N_{\mathrm{n}}} ) = \prod\limits_{i=1}^{N_{\mathrm{n}}} \mathscr{p}( \alpha_i \mid \mathrm{Pa}_{\alpha_i}^{\mathcal{G}} ) .
\end{align}
This factorization, known as the chain rule for \glspl{bn}, is more compact than the general chain rule for probabilities \citep[pp. 22, 51, 54, 62]{Koller2009}. The left-hand-side of the equation defines formally the joint \gls{pd} $\mathscr{p}(\cdot)$ of all \glspl{rv} shown in \figref{fig:bnet_of_stat_form}. The right-hand-side the conditional probabilities resulting from the graph structure. The event $\alpha_i$ presents a particular node $i$ of the \gls{dag} $\mathcal{G}$ with $N_{\mathrm{n}}=10$ nodes. The parent(s) of the $i$-th node are described by $\mathrm{Pa}_{\alpha_i}^{\mathcal{G}}$. Applying \eqref{eq:bn_chain_rule} on the \gls{dag} in \figref{fig:bnet_of_stat_form} we get the joint \gls{pd}  
\begin{align}
\label{eq:joint_prob}
\begin{aligned}
 &\mathscr{p}( \mathbf{X},\mathbf{Y},\mathbf{x},\mathbf{f},\sigma_{\delta}^2,\ell_{\delta},\hat{\mathbf{y}},\boldsymbol{\delta}_{\perp},\sigma_{\epsilon},\check{\mathbf{y}} ) = \\
 &\mathscr{p}( \mathbf{X} )\;
  \mathscr{p}( \mathbf{Y} )\;
  \mathscr{p}( \mathbf{f} )\;
  \mathscr{p}( \mathbf{x} )\;
  \mathscr{p}( \sigma_{\delta}^2 )\;
  \mathscr{p}( \ell_{\delta} )\;
  \mathscr{p}( \sigma_{\epsilon} )\;
  \mathscr{p}( \hat{\mathbf{y}} \mid \mathbf{X},\mathbf{Y},\mathbf{x} )\;
  \mathscr{p}( \boldsymbol{\delta}_{\perp} \mid \mathbf{f}, \mathbf{x},\sigma_{\delta}^2,\ell_{\delta} )\;
  \mathscr{p}( \check{\mathbf{y}} \mid \hat{\mathbf{y}},\boldsymbol{\delta}_{\perp},\sigma_{\epsilon} ) .
\end{aligned}
\end{align}
The \glspl{pd} $\mathscr{p}( \mathbf{X} )$, $\mathscr{p}( \mathbf{Y} )$ and $\mathscr{p}( \mathbf{f} )$ are omitted because they describe data already observed by simulation and are therefore fixed and represented by white circles in \figref{fig:bnet_of_stat_form}. By using the rule of conditional probability $\mathscr{p}( A,B )=\mathscr{p}( A \mid B )\; \mathscr{p}( B )$ \citep[p. 22]{Koller2009}, the left-hand-side of \eqref{eq:joint_prob} is represented as
\begin{align}
\begin{aligned}
  \mathscr{p}\big( \overbrace{(\hat{\mathbf{y}},\boldsymbol{\delta}_{\perp},\mathbf{x},\sigma_{\delta}^2,\ell_{\delta},\sigma_{\epsilon})}^{A},\overbrace{(\check{\mathbf{y}},\mathbf{X},\mathbf{Y},\mathbf{f})}^{B} \big) = \mathscr{p}\big( \overbrace{(\hat{\mathbf{y}},\boldsymbol{\delta}_{\perp},\mathbf{x},\sigma_{\delta}^2,\ell_{\delta},\sigma_{\epsilon})}^{A} \mid \overbrace{(\check{\mathbf{y}},\mathbf{X},\mathbf{Y},\mathbf{f})}^{B} \big)\; \mathscr{p}\big( \overbrace{(\check{\mathbf{y}},\mathbf{X},\mathbf{Y},\mathbf{f})}^{B} \big) 
\end{aligned}
\end{align}
and can be rewritten as
\begin{align}
\label{eq:complete_posterior}
\begin{aligned}
 &\mathscr{p}( \hat{\mathbf{y}},\boldsymbol{\delta}_{\perp},\mathbf{x},\sigma_{\delta}^2,\ell_{\delta},\sigma_{\epsilon} \mid \check{\mathbf{y}},\mathbf{X},\mathbf{Y},\mathbf{f} ) \\
 &=\frac{1}{\mathscr{p}( \check{\mathbf{y}},\mathbf{X},\mathbf{Y},\mathbf{f} )}
  \left[\mathscr{p}( \check{\mathbf{y}} \mid \hat{\mathbf{y}},\boldsymbol{\delta}_{\perp},\sigma_{\epsilon} )\;
  \mathscr{p}( \hat{\mathbf{y}} \mid \mathbf{X},\mathbf{Y},\mathbf{x} )\;
  \mathscr{p}( \boldsymbol{\delta}_{\perp} \mid \mathbf{f}, \mathbf{x},\sigma_{\delta}^2,\ell_{\delta} )\;
  \mathscr{p}( \mathbf{x} )\;
  \mathscr{p}( \sigma_{\delta}^2 )\;
  \mathscr{p}( \ell_{\delta} )\;
  \mathscr{p}( \sigma_{\epsilon} )\right] .
\end{aligned}
\end{align}
The left-hand-side of \eqref{eq:complete_posterior} describes the \gls{pd} known as the posterior distribution which describes the probability to observe the parameters $\boldsymbol{\chi} \coloneq [\mathbf{x}^\top, \sigma_{\delta}^2, \ell_{\delta}, \sigma_{\epsilon}]^\top$ and the quantities $\hat{\mathbf{y}}$ and $\boldsymbol{\delta}_{\perp}$ given our impedance measurements $\check{\mathbf{y}}$ and the obtained simulation data $\mathbf{X},\mathbf{Y},\mathbf{f}$. However, to infer exclusively the unknown parameters $\boldsymbol{\chi}$, we focus on a specific subset of variables in \eqref{eq:complete_posterior}. The quantities $\hat{\mathbf{y}}$ and $\boldsymbol{\delta}_{\perp}$ are thus regarded as nuisance parameters that complicate the inference of $\boldsymbol{\chi}$ \citep[pp.~328,~329]{Kay1993}. In Bayesian approaches, these parameters are removed by integrating them out \citep[pp.~328,~329]{Kay1993}. This process, known as marginalization, yields the marginalized posterior distribution 
\begin{align}
\label{eq:posterior_marginalization_concept}
\begin{aligned}
  & \mathscr{p}(\mathbf{x}, \sigma_{\delta}^2, \ell_{\delta}, \sigma_{\epsilon} \mid \check{\mathbf{y}},\mathbf{X},\mathbf{Y},\mathbf{f} ) = 
  \int_{\mathbb{R}^{M}} \int_{\mathbb{R}^{M}} \mathscr{p}( \hat{\mathbf{y}},\boldsymbol{\delta}_{\perp},\mathbf{x},\sigma_{\delta}^2,\ell_{\delta},\sigma_{\epsilon} \mid \check{\mathbf{y}},\mathbf{X},\mathbf{Y},\mathbf{f} ) \; \mathrm{d}\hat{\mathbf{y}}\; \mathrm{d}\boldsymbol{\delta}_{\perp} = \\
  &\frac{
  \overbrace{\left[ \int_{\mathbb{R}^{M}} \int_{\mathbb{R}^{M}} \mathscr{p}( \check{\mathbf{y}} \mid \hat{\mathbf{y}},\boldsymbol{\delta}_{\perp},\sigma_{\epsilon} )\;
  \mathscr{p}( \hat{\mathbf{y}} \mid \mathbf{X},\mathbf{Y},\mathbf{x} )\;
  \mathscr{p}( \boldsymbol{\delta}_{\perp} \mid \mathbf{f}, \mathbf{x},\sigma_{\delta}^2,\ell_{\delta} ) \; \mathrm{d}\hat{\mathbf{y}}\; \mathrm{d}\boldsymbol{\delta}_{\perp} \right]}^{\eqcolon\mathrm{I}}\;
  \mathscr{p}( \mathbf{x} )\;
  \mathscr{p}( \sigma_{\delta}^2 )\;
  \mathscr{p}( \ell_{\delta} )\;
  \mathscr{p}( \sigma_{\epsilon} ) 
  }{\mathscr{p}( \check{\mathbf{y}},\mathbf{X},\mathbf{Y},\mathbf{f} )}
\end{aligned}
\end{align}
which solves the inverse problem $\check{\mathbf{y}},\mathbf{X},\mathbf{Y},\mathbf{f} \mapsto \mathbf{x}, \sigma_{\delta}^2, \ell_{\delta}, \sigma_{\epsilon}$ in a probabilistic sense.
Substituting the data likelihood, the surrogate model predictive distribution, and the orthogonal discrepancy \gls{gp} prior from \eqref{eq:hierarchical_stat_formulation} into the marginalization integral $\mathrm{I}$ of \eqref{eq:posterior_marginalization_concept} yields 
\begin{align}
\mathrm{I} & =  \int_{\mathbb{R}^{M}} \left[ \int_{\mathbb{R}^{M}} \mathcal{N}\big( \check{\mathbf{y}} \mid \hat{\mathbf{y}} + \boldsymbol{\delta}_{\perp}, \sigma_{\epsilon}^2\; \mathbf{I} \big)\; \mathcal{N}\big( \hat{\mathbf{y}} \mid \boldsymbol{\mu}_{\hat{\mathbf{y}}}(\boldsymbol{\nu}_{\mathbf{x}}(\mathbf{x})), \boldsymbol{\Sigma}_{\hat{\mathbf{y}}}(\boldsymbol{\nu}_{\mathbf{x}}(\mathbf{x})) \big) \; \mathrm{d}\hat{\mathbf{y}} \right]\; \mathcal{N}\big( \boldsymbol{\delta}_{\perp} \mid \mathbf{0}, \mathbf{K}_{\boldsymbol{\delta},\perp}(\mathbf{x},\sigma_{\delta}^2, \ell_{\delta}) \big)\; \mathrm{d}\boldsymbol{\delta}_{\perp}  .
\end{align}
The integrand is a product of two nested marginal and conditional Gaussian distributions. Two successive applications of the marginalization rule for marginal and conditional Gaussians \citep[p.~92]{Bishop2006} yield the closed form
\begin{align}
\label{eq:res_marginalization_int}
\begin{aligned}
\mathrm{I} = \mathcal{N}\big( \check{\mathbf{y}} \mid \boldsymbol{\mu}_{\hat{\mathbf{y}}}(\boldsymbol{\nu}_{\mathbf{x}}(\mathbf{x})),\; \sigma_{\epsilon}^2\; \mathbf{I} + \boldsymbol{\Sigma}_{\hat{\mathbf{y}}}(\boldsymbol{\nu}_{\mathbf{x}}(\mathbf{x})) + \mathbf{K}_{\boldsymbol{\delta},\; \perp}(\mathbf{x},\sigma_{\delta}^2, \ell_{\delta}) \big) .
\end{aligned}
\end{align}
In \eqref{eq:res_marginalization_int} we summarize the \gls{iid} noise covariance and the two parametric covariances as  
\begin{align}
  \label{eq:combined_parametric_data_likelihood_cov}
  \boldsymbol{\Sigma}_{\check{\mathbf{y}}}(\mathbf{x},\sigma_{\delta}^2, \ell_{\delta}) \coloneq  \boldsymbol{\Sigma}_{\hat{\mathbf{y}}}(\boldsymbol{\nu}_{\mathbf{x}}(\mathbf{x})) + \mathbf{K}_{\boldsymbol{\delta},\perp}(\mathbf{x},\sigma_{\delta}^2, \ell_{\delta}) + \sigma_{\epsilon}^2 \; \mathbf{I}.
\end{align}
For the sake of numerical stability, \eqref{eq:combined_parametric_data_likelihood_cov} is regularized by
\begin{align}
  \label{eq:regularized_parametric_data_likelihood_cov}
  \boldsymbol{\Sigma}^{\prime}_{\check{\mathbf{y}}}(\mathbf{x},\sigma_{\delta}^2, \ell_{\delta}) \coloneq \boldsymbol{\Sigma}_{\check{\mathbf{y}}}(\mathbf{x},\sigma_{\delta}^2, \ell_{\delta}) + \epsilon_{\mathrm{rel}} \underbrace{\frac{\tr(\boldsymbol{\Sigma}_{\check{\mathbf{y}}}(\mathbf{x},\sigma_{\delta}^2, \ell_{\delta}))}{M}}_{\text{mean diagonal variance}}
\mathbf{I},
\end{align}
with relative jitter $\epsilon_{\mathrm{rel}}\coloneq 10^{-6}$. The three covariance terms in \eqref{eq:res_marginalization_int} are then replaced by \eqref{eq:regularized_parametric_data_likelihood_cov} and together with \eqref{eq:posterior_marginalization_concept} they form the final marginalized posterior distribution
\begin{align}
  \label{eq:marginalized_posterior}
  \mathscr{p}( \mathbf{x},\sigma_{\delta}^2,\ell_{\delta},\sigma_{\epsilon} \mid \check{\mathbf{y}},\mathbf{X},\mathbf{Y},\mathbf{f} ) = \dfrac{\mathcal{N}\big( \check{\mathbf{y}} \mid \boldsymbol{\mu}_{\hat{\mathbf{y}}}(\boldsymbol{\nu}_{\mathbf{x}}(\mathbf{x})), \boldsymbol{\Sigma}^{\prime}_{\check{\mathbf{y}}}(\mathbf{x},\sigma_{\delta}^2, \ell_{\delta}) \big) \; \mathscr{p}(\mathbf{x}) \; \mathscr{p}(\sigma_{\delta}^2) \; \mathscr{p}(\ell_{\delta}) \; \mathscr{p}(\sigma_{\epsilon})}{\mathscr{p}( \check{\mathbf{y}},\mathbf{X},\mathbf{Y},\mathbf{f} )},
\end{align}
which solves the inverse problem of our \gls{eesm} stator calibration task by providing us with a joint \gls{pd} for $\mathbf{x}, \sigma_{\delta}^2, \ell_{\delta}, \sigma_{\epsilon}$. It is solved using \gls{hmcmc} \citep{Brooks2011} with a \gls{nuts} \citep{Hoffman2014} implemented with Python's \emph{NumPyro} library \citep{Phan2019, Bingham2019}. This \gls{hmcmc}-based approach only requires the numerator of \eqref{eq:marginalized_posterior}. The denominator is a normalizing constant independent of the sampled parameters and thus drops out of the Metropolis(-Hastings) update step of \gls{hmcmc} \citep[pp.~122-124]{Brooks2011}.\par 
Since the geometry of the posterior in \eqref{eq:marginalized_posterior} is complex due to non-linear inter-parameter correlations (e.g., between $\mu_{\mathrm{r},0}$ and $f_{\mu}$ in \eqref{eq:first_order_debye_mdl}) and data likelihood inaccuracies from finite surrogate-model precision (cf. \secref{subsubsec:surrogate_accuracy}), we do not sample the unknown parameters $\mathbf{x}$, $\sigma_{\delta}^2$, $\ell_{\delta}$ and $\sigma_{\epsilon}$ directly. Instead, we use \gls{neutra} assisted \gls{nuts}-based \gls{hmcmc}, which applies a non-linear invertible transformation via an \gls{iaf} to sample in a less-complex transformed parameter space \citep{Hoffman2019}. The \emph{NumPyro} settings for the used \gls{neutra}-\gls{nuts}-\gls{hmcmc} are listed in \tabref{tab:neutra_nuts_settings}.
\begin{table}[!h]
  \centering
  \caption{\emph{NumPyro} \citep[version 0.22.0]{UAL2026} settings of the configured  Neural Transport Guide assisted NUTS-HMCMC.}
  \vspace{4pt}
  \label{tab:neutra_nuts_settings}
  {\scriptsize
  \begin{tabular}{ll | ll}
    \toprule
    \textbf{NUTS-based HMCMC} & & \textbf{Neural Transport Guide} & \\
    \midrule
    num\_warmup & 1000 & method & AutoIAFNormal \\
    num\_samples & 3000 & init\_loc\_fn & init\_to\_sample() \\
    num\_chains & 8 & num\_flows & 3 \\
    max\_tree\_depth & 12 & & \\
    dense\_mass & true & & \\
    target\_accept\_prob & 0.97 & & \\
    init\_strategy & init\_to\_sample() & & \\
    chain\_method & vectorized & & \\
    \bottomrule
  \end{tabular}
  }
\end{table}
The \gls{neutra} scheme uses a composition of three \glspl{iaf}. For \gls{nuts}-\gls{hmcmc}, we use $N_{\mathrm{ch}}\coloneq 8$ chains to explore multi-modality in the posterior \eqref{eq:marginalized_posterior}, which exceeds the minimum of 4 recommended by \citet[p. 671]{Vehtari2021} for multi-modality investigations. Each of the 8 \gls{nuts} chains is initialized randomly \citep[p.~673]{Vehtari2021} with 1000 warm-up iterations, where the target-probability of 0.97 is responsible for a fine step-size and mass-matrix adaptation in the \gls{hmcmc} scheme \citep[p.~38]{SDT2025_CmdStan}. The 1000 warm-up iterations are consistent with the default recommendations of the \citet[p.~333]{stan2025usersguide}. After warm-up, each \gls{nuts} chain draws $N_{\mathrm{draw}}\coloneq 3000$ post-warm-up samples and thereby constructs the posterior distribution in \eqref{eq:marginalized_posterior} with $N_{\mathrm{s}}\coloneq N_{\mathrm{ch}}\; N_{\mathrm{draw}} = \num{24e3}$ samples.\par
Convergence of the sampled posterior distribution is assessed using \gls{ess} and the rank-normalized $\widehat{\mathrm{R}}$ diagnostic \citep[p.~673]{Vehtari2021}. The \gls{ess} of the calibration parameters $\boldsymbol{\chi}$ is the number of independent samples that carry the same amount of information as the $N_{\mathrm{s}}$ dependent samples \citep[p.~2]{Eidsvik2020} obtained via \gls{hmcmc}. The higher the \gls{ess}, the more reliable the parameter estimates \citep[p.~672]{Vehtari2021}. The rank-normalized $\widehat{\mathrm{R}}$ diagnostic checks whether the $N_{\mathrm{ch}}$ chains have mixed well. It compares the variance of all the chains mixed together with the variance of individual chains. A variance ratio close to 1 indicates that all chains explore the same posterior \citep[p.~669]{Vehtari2021}. According to \citet[p. 95]{SDT2025_CmdStan} and \citet[p. 672]{Vehtari2021}, the \gls{ess} for all parameters should be "greater than 100 times the number of chains". The rank-normalized $\widehat{\mathrm{R}}$ value should be "smaller than~1.01" \citep[p. 671]{Vehtari2021}.\par
If both convergence conditions are satisfied for all calibration parameters $\boldsymbol{\chi}$, statistical values derived from the posterior samples are reliable. For instance, we can compute the "mean" (first moment)
\begin{align}
  \label{eq:posterior_mean}
  \boldsymbol{\chi}_{\mathrm{mean}} = [\mathbf{x}_{\mathrm{mean}}^\top, \boldsymbol{\theta}_{\delta,\mathrm{mean}}^\top, \sigma_{\epsilon,\mathrm{mean}} ]^\top \coloneq \expect\left[ \boldsymbol{\chi} \mid \check{\mathbf{y}},\mathbf{X},\mathbf{Y},\mathbf{f} \right]
\end{align}
with the expectation operator $\expect[\cdot]$, as well as the maximum a posteriori "map" estimate
\begin{align}
  \label{eq:posterior_map}
  \boldsymbol{\chi}_{\mathrm{map}} = [\mathbf{x}_{\mathrm{map}}^\top, \boldsymbol{\theta}_{\delta,\mathrm{map}}^\top, \sigma_{\epsilon,\mathrm{map}}]^\top \coloneq \argmax\limits_{\boldsymbol{\chi}} \mathscr{p}(\boldsymbol{\chi} \mid \check{\mathbf{y}},\mathbf{X},\mathbf{Y},\mathbf{f} ) .
\end{align}
While variance or credible intervals can quantify output variability, we primarily aim for a well-calibrated simulation model \eqref{eq:sim_mdl} and therefore do not address other statistics further.

%% file: 04_main_content/04_bayesian_inference_results.tex
\section{Bayesian Calibration Results and Discussion}
\label{sec:bayesian_cal_results}
In this section we apply the probabilistic \gls{emp} calibration strategy developed in \secref{sec:cal_problem} and \secref{sec:bayesian_inf_formulation} to our \gls{eesm} stator. It is tested on the stator \gls{dm} impedance, where the goal is the calibration of the \gls{dm} simulation model~\eqref{eq:sim_mdl} against the measured \gls{dm} impedance. We seek the \glspl{emp} $\mathbf{x}$ that minimize the deviation between the 3D CST \gls{fe} simulated impedance $\mathbf{y} = \boldsymbol{\mathcal{M}}(\mathbf{x})$ and the observed impedance $\check{\mathbf{y}}$. To this end, we examine and discuss three calibration scenarios that differ in the prior ranges of the \glspl{emp}.\par
The \gls{emp} ranges of the three investigated calibration scenarios are listed in \tabref{tab:inference_cases}. In scenario (a), we aim to calibrate all four \glspl{emp} $\mathbf{x}$ across the entire surrogate input-range (cf. \tabref{tab:design_parameter_ranges}).
\begin{table}[!h]
  \caption{Parameter ranges defining the material parameters' LogUniform prior distributions for three calibration scenarios of the EESM-stator.}
  \vspace{4pt}
  \label{tab:inference_cases}
  \renewcommand{\arraystretch}{1.0}
  \centering
  {\scriptsize
  \begin{tabularx}{\linewidth}{
      l|
      c|
      c|
      >{\centering\arraybackslash}X|
      >{\centering\arraybackslash}X|
      >{\centering\arraybackslash}X|
      >{\centering\arraybackslash}X|
      >{\centering\arraybackslash}X|
      >{\centering\arraybackslash}X
    }
    \toprule
    \multicolumn{1}{c}{} & \multicolumn{1}{c}{} & & \multicolumn{2}{c|}{\makecell{full calibration with \\ EMP ranges reduced \\ to main-mode}} & \multicolumn{2}{c|}{\makecell{calibration without \\ permeability and entire \\ surrogate ranges for the EMPs}} & \multicolumn{2}{c}{\makecell{full calibration with \\ ``datasheet-aligned'' \\ EMP ranges}} \\
    \multicolumn{1}{c}{} & \multicolumn{1}{c}{} & & \multicolumn{2}{c|}{} & \multicolumn{2}{c|}{} & \multicolumn{2}{c}{} \\
    \multicolumn{1}{c}{} & \multicolumn{1}{c}{} & & \multicolumn{2}{c|}{scenario (a)} & \multicolumn{2}{c|}{scenario (b)} & \multicolumn{2}{c}{scenario (c)} \\
    \midrule
     $d$ & Param. & Unit & $x_{d}^{\mathrm{l}}$ & $x_{d}^{\mathrm{u}}$ & $x_{d}^{\mathrm{l}}$ & $x_{d}^{\mathrm{u}}$ & $x_{d}^{\mathrm{l}}$ & $x_{d}^{\mathrm{u}}$ \\
    \midrule
    1 & $\mu_{\mathrm{r,0}}$ & - & \num{8e2} & \num{4e3} & --- & --- & \num{1e2} & \num{4e3} \\
    2 & $f_{\mu}$ & $\si{\hertz}$ & \num{1e4} & \num{700e3} & \num{8e2} & \num{1.25e6} & \num{1e4} & \num{700e3} \\
    3 & $\varepsilon_{\mathrm{r}}$ & - & 1 & 12 & \num{0.75} & \num{12} & 1 & 12 \\
    4 & $\sigma_{\mathrm{e},\parallel}$ & $\si{\siemens/\meter}$ & \num{4100} & \num{7e3} & \num{1e2} & \num{1e7} & \num{1e5} & \num{1e7} \\
    \bottomrule
  \end{tabularx}
  }
\end{table}
However, we found that this leads to a posterior distribution with multiple modes. This happens because part of the magnetic field around a stator hairpin closes through the air between the stator teeth (cf. \figref{fig:3d_cst_stator_geometry}). Therefore, this air gap primarily defines the stator inductance. Because the stator inductance and subsequently the \gls{dm} impedance are relatively insensitive to changes in relative permeability, multi-modality occurs. For this reason, we reduced the \gls{emp} ranges in scenario (a) to values around an initially identified posterior main-mode. In scenario (b), we use the full surrogate input-range but fix the relative permeability to $\mu_{\mathrm{r,0}}\coloneq 870$, following NO25 material values \citep{VoestalpineNO25}. In scenario (c), the \gls{emp} ranges are narrowed to values close to those specified for the used NO25 material.\par
\begin{table}[!b]
\centering
\caption{Summary of the Bayesian calibration statistics (mean, map) and diagnostics (ESS, $\widehat{\mathrm{R}}$) for three calibration scenarios using a NeuTra-assisted NUTS-HMCMC.}
\vspace{4pt}
\label{tab:summary_of_mcmc_inference_cases}
{\scriptsize
\begin{tabularx}{\linewidth}{
l|
>{\centering\arraybackslash}X
>{\centering\arraybackslash}X
>{\centering\arraybackslash}X
>{\centering\arraybackslash}X|
>{\centering\arraybackslash}X
>{\centering\arraybackslash}X
>{\centering\arraybackslash}X
>{\centering\arraybackslash}X|
>{\centering\arraybackslash}X
>{\centering\arraybackslash}X
>{\centering\arraybackslash}X
>{\centering\arraybackslash}X
}
\toprule
 & \multicolumn{4}{c|}{scenario (a)} & \multicolumn{4}{c|}{scenario (b)} & \multicolumn{4}{c}{scenario (c)} \\
\midrule
 & map & mean & ESS & $\widehat{\mathrm{R}}$ & map & mean & ESS & $\widehat{\mathrm{R}}$ & map & mean & ESS & $\widehat{\mathrm{R}}$ \\
\midrule
$\mu_{\mathrm{r,0}}$ & \num{1.99e3} & \num{2.01e3} & \num{20262} & \num{1.000} & --- & --- & --- & --- & \num{3.77e3} & \num{3.82e3} & \num{25245} & \num{1.000} \\
\midrule
$f_{\mu}$ & \SI{178}{\kilo\hertz} & \SI{179}{\kilo\hertz} & \num{18931} & \num{1.000} & \SI{147}{\kilo\hertz} & \SI{145}{\kilo\hertz} & \num{19640} & \num{1.000} & \SI{108}{\kilo\hertz} & \SI{102}{\kilo\hertz} & \num{17107} & \num{1.000} \\
\midrule
$\varepsilon_{\mathrm{r}}$ & \num{3.998} & \num{3.993} & \num{15248} & \num{1.000} & \num{4.166} & \num{4.163} & \num{18147} & \num{1.000} & \num{4.801} & \num{4.816} & \num{12906} & \num{1.000} \\
\midrule
$\sigma_{\mathrm{e},\parallel}$ & \SI{4.26e3}{\siemens/\meter} & \SI{4.27e3}{\siemens/\meter} & \num{21247} & \num{1.000} & \SI{2.59e3}{\siemens/\meter} & \SI{2.57e3}{\siemens/\meter} & \num{13783} & \num{1.000} & \SI{1.02e5}{\siemens/\meter} & \SI{1.01e5}{\siemens/\meter} & \num{22327} & \num{1.000} \\
\midrule
$\sigma_{\delta}^2$ & \num{7.15e-3} & \num{6.99e-3} & \num{19612} & \num{1.000} & \num{8.96e-3} & \num{9.21e-3} & \num{16910} & \num{1.000} & \num{1.37e-2} & \num{4.38e-2} & \num{11932} & \num{1.001} \\
\midrule
$\ell_{\delta}$ & \num{1.65e-2} & \num{1.60e-2} & \num{19734} & \num{1.001} & \num{1.79e-2} & \num{1.81e-2} & \num{20473} & \num{1.000} & \num{6.88e-2} & \num{9.93e-2} & \num{6058} & \num{1.001} \\
\midrule
$\sigma_{\epsilon}$ & \num{6.22e-4} & \num{5.24e-4} & \num{21604} & \num{1.000} & \num{6.38e-4} & \num{4.57e-4} & \num{19584} & \num{1.000} & \num{1.25e-1} & \num{1.24e-1} & \num{17281} & \num{1.000} \\
\bottomrule
\end{tabularx}
}
\vspace{-1\baselineskip}
\end{table}
\tabref{tab:summary_of_mcmc_inference_cases} summarizes the posterior "mean" \eqref{eq:posterior_mean} and "map" \eqref{eq:posterior_map} values, along with the convergence diagnostics (\gls{ess} and rank-normalized $\widehat{\mathrm{R}}$), for all parameters $\boldsymbol{\chi}$ across the three scenarios after running the \gls{neutra}-\gls{nuts}-\gls{hmcmc}. The total computation time for each scenario is less than $\SI{20}{\minute}$ on an NVIDIA A100 GPU. For all three scenarios all convergence conditions are met, with \gls{ess} larger than 800 (100 times the number of \gls{hmcmc} chains) and rank-normalized $\widehat{\mathrm{R}}$ smaller than 1.01. Given that the convergence conditions are met in all three scenarios, the "mean" and "map" estimates of $\boldsymbol{\chi}$ are reliable. The most consistent quantity is the equivalent relative permittivity $\varepsilon_{\mathrm{r}}$ of the hairpin coating. Its "mean" and "map" values lie between 4.0 and 4.8 across all scenarios. However, this similarity is not given for the other calibration parameters $\boldsymbol{\chi} \setminus \varepsilon_{\mathrm{r}}$ showing a significantly larger spread of values across all scenarios.\par
\begin{figure}[!t]
  \centering
  \includegraphics[scale=0.98]{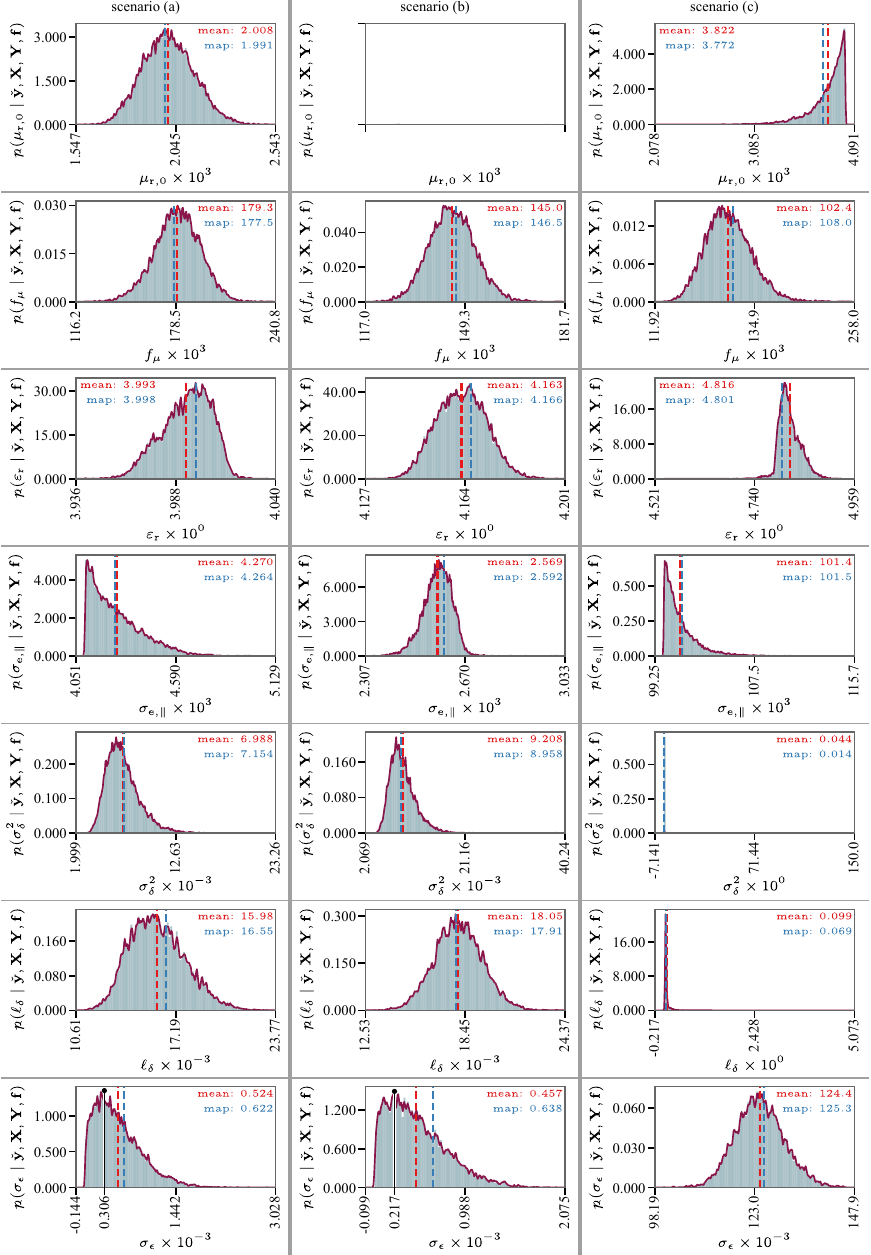}
  \caption{Marginal posterior distributions $\mathscr{p}( \; \cdot \mid \check{\mathbf{y}},\mathbf{X},\mathbf{Y},\mathbf{f} )$ of the EESM-stator calibration problem obtained using NeuTra-assisted NUTS-HMCMC for three calibration scenarios.}
  \label{fig:marginal_posterior_dists}
\end{figure}
\begin{figure}[!b]
    \centering
     \begin{minipage}[b]{\textwidth}
        \centering
        \includegraphics[scale=0.995]{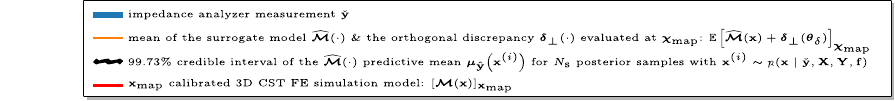}
    \end{minipage}   
    \begin{minipage}[b]{\textwidth}
        \centering
        \includegraphics[scale=0.995]{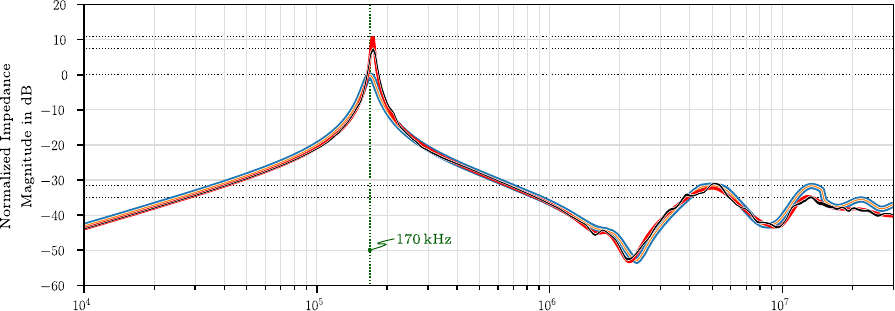}
        \subcaption{}
        \vspace{3mm}
        \label{fig:meas_surrogate_cst_impedance_scen_a}
    \end{minipage}
    \begin{minipage}[b]{\textwidth}
        \centering
        \includegraphics[scale=0.995]{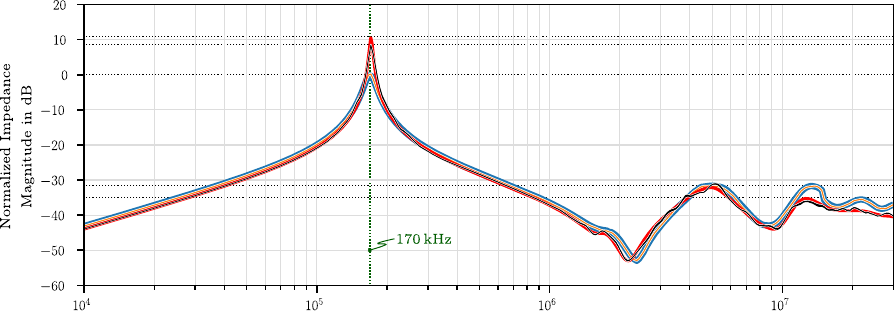}
        \subcaption{}
        \vspace{3mm}
        \label{fig:meas_surrogate_cst_impedance_scen_b}
    \end{minipage}
    \begin{minipage}[b]{\textwidth}
        \centering
        \includegraphics[scale=0.995]{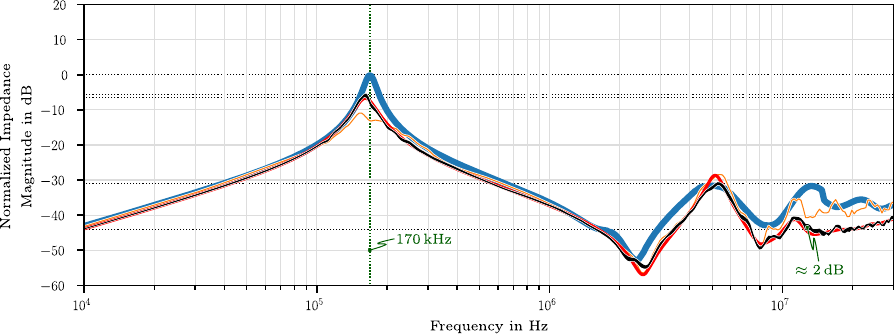}
        \subcaption{}
        \label{fig:meas_surrogate_cst_impedance_scen_c}
    \end{minipage}
    \caption{Comparison of the measured (blue) EESM-stator DM impedance against the surrogate model predictive mean \eqref{eq:predictive_mu} and its variation across the posterior samples (black), the surrogate and discrepancy model based prediction (orange), and the 3D CST FE simulation (red) for the three calibration scenarios. The orange and red traces are evaluated at the maximum a posteriori "map" parameters of each scenario.}
    \label{fig:meas_surrogate_cst_impedance}
\end{figure}
Therefore, visualizing the marginal posterior distributions resulting from \eqref{eq:marginalized_posterior} in \figref{fig:marginal_posterior_dists} provides further insight. \figref{fig:marginal_posterior_dists} consists of $7\times 3$ subplots, one for each marginalized posterior distribution. The three columns correspond to the three calibration scenarios, and the seven rows to the individual parameter entries of $\boldsymbol{\chi}$. Each subplot shows the \gls{pd} of \eqref{eq:marginalized_posterior} marginalized with respect to one parameter, i.e., the \gls{pd} of that parameter given the \gls{dm} measurement $\check{\mathbf{y}}$, the underlying surrogate dataset $\{\mathbf{X},\mathbf{Y}\}$ and observation frequencies $\mathbf{f}$. The red and blue dashed lines mark the "mean" and the "map" of each marginal distribution, respectively. Both values are previously tabulated in \tabref{tab:summary_of_mcmc_inference_cases}.\par
Starting at the top of \figref{fig:marginal_posterior_dists} with the static permeability $\mu_{\mathrm{r,0}}$ of the first-order Debye model. Scenario~(a) shows a symmetric, Gaussian-like distribution with nearly identical "mean" and "map" values of about 2,000. In scenario~(b), there is no distribution for $\mu_{\mathrm{r,0}}$. We fixed $\mu_{\mathrm{r,0}}$ by definition to analyze the previously mentioned air-gap-related permeability identification problem. In scenario (c), the distribution differs significantly from (a). It is neither symmetric nor Gaussian-like, but strongly right-skewed. The most probable permeability values lie at the edge of the defined scenario parameter range, i.e., at the boundary of the surrogate model's input range (cf. \tabref{tab:design_parameter_ranges}), with permeability values close to 4,000. Moreover, enforcing conductivity values $\sigma_{\mathrm{e},\parallel}>\SI{1e5}{\siemens\per\meter}$ in scenario (c) drives the permeability toward larger values and the conductivity toward smaller values. The skewness of both distributions results in marginal distributions of $\mu_{\mathrm{r,0}}$ and $\sigma_{\mathrm{e},\parallel}$ that both lie at the edge of the scenario parameter range. This is also visible in the corresponding "mean" and "map" values of $\mu_{\mathrm{r,0}}$ and $\sigma_{\mathrm{e},\parallel}$ of approximately $\num{3.8e3}$ and $\SI{101e3}{\siemens\per\meter}$, respectively. Compared with (a) and (c), the $\sigma_{\mathrm{e},\parallel}$ distribution of scenario (b) is much more symmetric and Gaussian-like, with "mean" and "map" values around $\num{2.6e3}$. The relaxation frequency $f_{\mu}$ is the last of the four material parameters $\mathbf{x}$ to be calibrated. In all three cases, the distributions are very symmetric and Gaussian-like. Their "mean" and "map" values are approximately $\SI{180}{\kilo\hertz}$ (scenario (a)), $\SI{145}{\kilo\hertz}$ (scenario (b)), and $\SI{105}{\kilo\hertz}$ (scenario (c)). All four \glspl{emp} share the same behavior. Based on the known measurement $\check{\mathbf{y}}$ and simulation dataset $\{\mathbf{X},\mathbf{Y},\mathbf{f}\}$, the initially assumed LogUniform prior distributions are updated using \eqref{eq:marginalized_posterior}. The resulting distributions are more Gaussian-like. The most probable \gls{emp} ranges are significantly reduced. This can be seen by comparing the x-axes of the first 4 rows of \figref{fig:marginal_posterior_dists} with the \gls{emp} ranges in \tabref{tab:inference_cases}.\par
Besides the four \glspl{emp}, \figref{fig:marginal_posterior_dists} also shows the marginal distributions of the orthogonal discrepancy hyperparameters $\boldsymbol{\theta}_{\delta} = [\sigma_{\delta}^2 , \ell_{\delta}]^\top$, as well as the marginal distribution of the noise standard deviation $\sigma_{\epsilon}$. In scenarios (a) and (b), the distributions of $\sigma_{\delta}^2$ and $\ell_{\delta}$ are relatively symmetric and Gaussian-like. Compared with the needle-like distributions in~(c), they have much shorter right-sided tails. In scenarios (a) and (b), the "mean" and "map" of each distribution are nearly identical. In scenario (c), the "mean" and "map" values of $\sigma_{\delta}^2$ and $\ell_{\delta}$ differ strongly. The "mean" value of $\ell_{\delta}$ is about 150\% larger and that of $\sigma_{\delta}^2$ about 320\% larger than the respective "map" values. The noise standard deviation $\sigma_{\epsilon}$ shows the opposite behavior with respect to symmetry. In scenarios (a) and (b), the shape of the initially imposed Beta distribution is largely preserved. However, the mode of the distribution (black vertical stem marker) moves to values 30 to 45 times smaller than the initially assumed mode of 0.01 (cf. \secref{sec:bayesian_inf_formulation}). Scenario (c) behaves differently. Here, the initial Beta distribution updates to a Gaussian-like distribution. The "mean" and "map" values in (c) are approximately 200 to 250 times larger than in scenarios (a) and (b). This 200 to 250 times larger noise standard deviation, together with the 4 to 6 times larger length-scale $\ell_{\delta}$ in (c) compared to (a) and (b), and the fact that both $\mu_{\mathrm{r,0}}$ and $\sigma_{\mathrm{e},\parallel}$ distributions lie at the edge of the scenario parameter range, indicates that scenario (c) cannot be well calibrated with the \glspl{emp} alone. Correction must instead be provided by the noise and the orthogonal discrepancy term.\par
This is confirmed by evaluating the predictive mean $\boldsymbol{\mu}_{\hat{\mathbf{y}}}(\mathbf{x}^{(i)})$ of the \gls{wbk} surrogate model for all $N_{\mathrm{s}}=\num{24e3}$ posterior samples $\mathbf{x}^{(i)} \sim \mathscr{p}(\mathbf{x} \mid \check{\mathbf{y}},\mathbf{X},\mathbf{Y},\mathbf{f})$ and comparing it with the measurement $\check{\mathbf{y}}$, as shown in \figref{fig:meas_surrogate_cst_impedance}. The black spanned region describes the variation range of the \gls{eesm}-stator \gls{dm} impedance caused by 99.73\% of all posterior samples (\gls{wbk} surrogate model is evaluated for the posterior samples). If the variation were Gaussian, this would correspond to a 6-sigma band. Evaluating the surrogate model for all posterior samples results in an uncertainty quantification of the \gls{dm} impedance. For all scenarios (\figref{fig:meas_surrogate_cst_impedance_scen_a} to \figref{fig:meas_surrogate_cst_impedance_scen_c}), the spanned black variation band is about 2\,dB in the worst case (\figref{fig:meas_surrogate_cst_impedance_scen_c}). In \figref{fig:meas_surrogate_cst_impedance_scen_a} and \figref{fig:meas_surrogate_cst_impedance_scen_b}, the black variation band matches the blue measurement well over the whole frequency range, although local deviations exist. At the main resonance of about $\SI{170}{\kilo\hertz}$, both scenarios show a magnitude error of about 8\,dB. For frequencies above $\SI{10}{\mega\hertz}$, deviations of up to $\pm 5\,\text{dB}$ occur in both cases. No parameter combination from the posterior distribution can correct these deviations. This holds even more for scenario (c) in \figref{fig:meas_surrogate_cst_impedance_scen_c}. Compared with the measurement, the absolute maximum deviations of the black variation band remain roughly the same. However, the broadband agreement is worse due to a shift in the impedance characteristic. Since the black variation band is very narrow for all three scenarios, we calibrate the 3D CST \gls{fe} simulation model $\boldsymbol{\mathcal{M}}(\mathbf{x})$ with a single \gls{emp} vector per scenario. This \gls{emp} vector is defined by the "map" values, which represent the most probable values. The calibrated simulation model $\boldsymbol{\mathcal{M}}(\mathbf{x}_{\mathrm{map}})$ is shown by the red curves in \figref{fig:meas_surrogate_cst_impedance_scen_a} to \figref{fig:meas_surrogate_cst_impedance_scen_c}. It behaves similarly to the black variation band, but with fewer local fluctuations and a slightly larger absolute magnitude error of about 10\,dB at $\SI{170}{\kilo\hertz}$. This is likely due to the finite accuracy of the surrogate. It is evident that the differences between the measurement and the surrogate model, and between the measurement and the simulation model, cannot be eliminated by a suitable choice of \glspl{emp} $\mathbf{x}$. There is thus a certain model discrepancy between the measurement $\check{\mathbf{y}}$ and the simulation model $\boldsymbol{\mathcal{M}}(\mathbf{x})$. It can be described by a discrepancy term orthogonal to the surrogate (and simulation) model, as shown by the orange curve in \figref{fig:meas_surrogate_cst_impedance_scen_a} to \figref{fig:meas_surrogate_cst_impedance_scen_c}. The orange curve is the superposition of the predictive mean of the surrogate model and the mean of the discrepancy term, both evaluated at "map" values. This yields perfect agreement with the measurement in \figref{fig:meas_surrogate_cst_impedance_scen_a} and \figref{fig:meas_surrogate_cst_impedance_scen_b}. However, not for the scenario in \figref{fig:meas_surrogate_cst_impedance_scen_c}. Here, the requirement of conductivity values $\sigma_{\mathrm{e},\parallel}>\SI{1e5}{\siemens\per\meter}$ is a case where even the orthogonal discrepancy term lacks the flexibility to correct all impedance differences. This is visible in the differences between the orange and blue curves.\par
In summary, while the \gls{fe} simulation model \eqref{eq:sim_mdl} can be calibrated to an accuracy of 10\,dB, an exact calibration is achievable by superposing the corresponding surrogate model \eqref{eq:emu_mdl} with the orthogonal discrepancy term \eqref{eq:ortho_delta}. Two cases must be distinguished. Case 1 concerns scenarios (a) and (b), where the \gls{dm} impedance is well represented over the whole frequency range, apart from local inaccuracies. These results are achievable with "datasheet-deviating" parameter values. Case 2 corresponds to scenario (c), which performs worse, even though the calibration uses "datasheet-aligned" values. For Case 1, the orthogonal discrepancy term corrects the residual deviations where the underlying model contributes little. In scenarios (a) and (b), the permeability values are consistent with the used NO25 lamination material \citet{VoestalpineNO25}. The relaxation frequencies are consistent with the impedance point of highest damping at $\SI{170}{\kilo\hertz}$. The material conductivities, however, are too low. This suggests that the frequency-dependent losses of the magnetic core behavior (cf. \eqref{eq:complex_permeability}) cannot be perfectly captured by the chosen first-order Debye model. Consequently, it must be compensated by a reduced sheet conductivity $\sigma_{\mathrm{e},\parallel}$, denoted as "datasheet-deviating". This is because, from an impedance perspective, both the sheet conductivity~$\sigma_{\mathrm{e},\parallel}$~\eqref{eq:conductivity_tensor} and the frequency-dependent losses $\mu_{\mathrm{r}}^{\prime\prime}(f;\mu_{\mathrm{r},0},\mu_{\mathrm{r},\infty},f_{\mu})$ \eqref{eq:complex_permeability} qualitatively produce the same resistive effect. While $\sigma_{\mathrm{e},\parallel}$ provides a constant impedance correction, $\mu_{\mathrm{r}}^{\prime\prime}(f;\cdot)$ introduces a frequency-dependent correction. Therefore, they can mutually compensate for one another to adjust the overall resistive \gls{dm} impedance component. Although the resulting conductivity is not physically interpretable, the "datasheet-deviating" values still yield a satisfactory impedance calibration result from an \gls{emc} perspective. This confirms the model's practical usefulness and reflects the pragmatic view of Box and Draper: \emph{``Remember that all models are wrong; the practical question is how wrong do they have to be to not be useful.''} \citep[p.~74]{Box1987}. Therefore, the calibrated "map" parameter sets of scenarios (a) or (b) serve as potential initial values for more elaborate \gls{emc} system-level simulations, e.g., of an entire \gls{ev} powertrain at previously inaccessible frequencies up to $\SI{30}{\mega\hertz}$, in which the electrical machine forms a major part of the \gls{emi} propagation path.

%% file: 04_main_content/05_conclusion.tex
\section{Conclusion}
%
%
In this article, we developed a surrogate model based material parameter calibration strategy for an \gls{eesm} stator based on a Bayesian model calibration formation which accounts for model discrepancies. Its goal is twofold: to identify the \glspl{emp} that best reproduce the measured broadband \gls{dm} impedance up to $\SI{30}{\mega\hertz}$, and to determine whether the underlying material modeling is adequate.\par
%
%
Based on three examined calibration scenarios, we found that \glspl{emp} with "datasheet-deviating" values yield a good broadband impedance calibration. The normalized impedance deviations remain below $\pm 5\,\text{dB}$ up to $\SI{30}{\mega\hertz}$, except at the \gls{dm} impedance's main resonance of $\SI{170}{\kilo\hertz}$, where they reach about $10\,\text{dB}$. Together with the introduced model discrepancy term, we found an inadequacy in the material modeling. The first-order Debye model cannot fully capture the frequency-dependent losses of the \gls{eesm} stator, which gives rise to the "datasheet-deviating" \glspl{emp}.\par
%
%
The proposed calibration methodology enables us to estimate unknown material properties from impedance measurements. Although the identified "datasheet-deviating" \glspl{emp} may differ from the values reported in datasheets, they can still be used to estimate characteristics such as the frequency-dependent impedance of an electrical machine, which is particularly relevant for subsequent analyses, including \gls{emc} simulations and \gls{emi} prediction.\par
An advantage of the proposed approach is that it does only require little prior knowledge of the material properties, making it a useful tool for obtaining initial estimates for further investigations. Moreover, the methodology can be extended to more complex problems, such as the calibration of \glspl{emp} for an entire electrical machine.\par
\section{Outlook}
%
%
Three directions for improvement are envisioned for future work. First, alternative material modeling approaches could be explored, such as the lamination-aware permeability modeling demonstrated by \citet[pp.~13--15]{Jorks2015}, \citet{Hewitt2024}, \citet{Shheibar2026}, and \citet[pp. 176, 178, eq.~A.11]{Meeker2026}. Alternatively, one could investigate a generalized Debye relaxation model ---~also known as Havriliak-Negami process \citep[p.~162]{Havriliak1967}~--- which provides two additional degrees of freedom over the used first-order Debye model. This could potentially reduce the modeling inadequacies caused by the used first-order Debye model and thus result in \glspl{emp} with "datasheet-aligned" values. Second, the surrogate model accuracy can be improved to yield less complex posterior distributions. Third, extending the approach to higher-dimensional calibrations with $D > 4$ \glspl{emp}.

%% file: 05_appendix/01_appendix.tex
%

%% file: 06_acknowledgement/01_acknowledgement.tex
\section*{Acknowledgement}
\label{sec:acknowledgement}
%
%
%
This work was supported in part by the Austrian Federal Ministry for Digital and Economic Affairs, the National Foundation for Research, Technology and Development and the Christian Doppler Research Association and in part by Silicon Austria Labs (SAL) GmbH, owned by the Republic of Austria, the Styrian Business Promotion Agency (SFG), the Federal State of Carinthia, the Upper Austrian Research (UAR), and the Austrian Association for the Electric and Electronics Industry (FEEI).